# Composite antiferromagnetic and orbital order with altermagnetic properties at a cuprate/manganite interface

**Subhrangsu Sarkar**[a,1], **Roxana Capu**[b,1], **Yurii G. Pashkevich**[a,c], **Jonas Knobel**[a], **Marli R. Cantarino**[a,d], **Abhishek Nag**[e], **Kurt Kummer**[d], **Davide Betto**[d], **Roberto Sant**[d], **Christopher W. Nicholson**[a], **Jarji Khmaladze**[a], **Ke-jin. Zhou**[e], **Nicholas B. Brookes**[d], **Claude Monney**[a], **and Christian Bernhard**[a,1]

**Heterostructures from complex oxides allow one to combine various electronic and magnetic orders as to induce new quantum states. A prominent example is the coupling between superconducting and magnetic orders in multilayers from high-Tc cuprates and manganites. A key role is played here by the interfacial $CuO_2$ layer whose distinct properties remain to be fully understood. Here, we study with resonant inelastic X-ray scattering (RIXS) the magnon excitations of this interfacial $CuO_2$ layer. In particular, we show that the underlying antiferromagnetic exchange interaction at the interface is strongly suppressed to $J \approx 70$ meV, as compared to $J \approx 130$ meV for the $CuO_2$ layers away from the interface. Moreover, we observe an anomalous momentum dependence of the intensity of the interfacial magnon mode and show that it suggests that the antiferromagnetic order is accompanied by a particular kind of orbital order that yields a so-called altermagnetic state. Such a two-dimensional altermagnet has recently been predicted to enable new spintronic applications and superconducting proximity effects.**

## Significance Statement

We report a resonant inelastic X-ray scattering study of multilayers made from a cuprate high-Tc superconductor and a magnetic perovskite manganite. Our study reveals an extraordinary behavior of the spins and electrons of the interfacial cuprate monolayer. In particular, we observe that its antiferromagnetic spin interaction is strongly suppressed and we make the fundamental discovery that it hosts a new kind of combined magnetic and electronic order that constitutes a two-dimensional altermagnetic state. Our findings significantly advance the state of the art in the field of altermagnets that are of great current interest since they enable new kinds of spintronic and magnonic devices and may also lead to exotic superconducting proximity effects.

The proximity effect at the interfaces between different materials with strongly correlated electrons, such as Mott-type antiferromagnetic (AF) insulators, cuprate high-$T_c$ superconductors or the colossal-magneto-resistance (CMR) manganites, is of great current interest in fundamental and applied since. Specifically, the cuprate/manganite interface holds great promises for inducing unconventional quantum states which have potential applications e.g. in spintronics or quantum computation.

This creates an urgent need for experimental techniques which can selectively probe the electronic and magnetic properties in the vicinity of the interfaces of such heterostructures. In the following, we demonstrate for the case of a cuprate/manganite interface that resonant inelastic X-ray scattering (RIXS) can provide unique information about the antiferromagnetic exchange interaction and the orbital order of the holes on the interfacial $CuO_2$ layer.

The parent compounds of the high-Tc cuprates are charge transfer insulators with a long-range AF order of the spins of the holes that reside on half-filled Cu-$3d_{x^2-y^2}$ levels. Upon doping away from half-filling (the additional holes have a strong oxygen character and form so-called Zhang-Rice singlets (1)), the long-range AF order is rapidly suppressed and a strongly-correlated conducting and eventually superconducting state develops. Static but short ranged AF correlations persist to a higher hole doping and coexist with the superconducting (SC) order in parts of the so-called underdoped regime where $T_c$ increases with doping and eventually reaches a maximum around optimal doping.

Fluctuating AF correlations persist even beyond optimum doping (2) into the so-called overdoped regime where SC is suppressed and eventually vanishes. It is therefore widely assumed that the AF fluctuations are involved in the SC pairing (3). However, a consensus on this issue has not yet been reached, since the strong electronic correlations also give rise to a short-ranged charge order(4–7), and possibly even quadrupolar or octopolar orders(8) or so-called flux-phases (9), that may also coexist with high-Tc superconductivity(10).

The properties of the AF spin fluctuations and their evolution upon hole doping have been extensively investigated with inelastic neutron scattering (INS). The studied materials range from $La_2CuO_4$ (11–13), to $YBa_2Cu_3O_{6+\delta}$ (13–16), $La_{2-x}(Ba,Sr)_xCuO_4$(13, 17) and other cuprates like Bi-2212 and Hg-1201(18).

Author affiliations: [a]University of Fribourg, Department of Physics and Fribourg Center for Nanomaterials, Chemin du Musée 3, CH-1700 Fribourg, Switzerland; [b]West University of Timisoara, Department of Physics, Bd Vasile Parvan 4, Timisoara-300223, Romania; [c]O. Galkin Donetsk Institute for Physics and Engineering NAS of Ukraine, 03028 Kyiv, Ukraine; [d]European Synchrotron Radiation Facility, 71 Avenue des Martyrs, CS40220, F-38043 Grenoble Cedex 9, France; [e]Diamond Light Source, Harwell Campus, Didcot, Oxfordshire OX11 0DE, United Kingdom

---

The sample was prepared by J. K. The RIXS experiments were performed by S.S, R.C, M.C, C.W.M and C.M with onsite assistance from A.N, K.Z, D.B, K.K and N.B. The XAS measurements were performed by S.S, R.C, C.B and J. K. with the onsite help from R.S, D.B, K.K and N.B. S.S and R.C analyzed the data with the help from C.M, A.N and K.K. The theoretical modeling was performed by Y.P. The manuscript was prepared by S.S, C.B, Y.P, and R.C with critical suggestions from C.M, C.W.M, A.N, N.B and R.S. J.K helped in LaTeX compilation.

The authors declare no competing interest.

[1]To whom correspondence should be addressed. E-mail: subhrangsu.sarkar@unifr.ch, roxana.capu@e-uvt.ro, christian.bernhard@unifr.ch

The INS studies have shown that the AF exchange interaction is strongly anisotropic, i.e. for YBCO the in-plane interaction amounts to $J_{\parallel} \approx 120-130$ meV, whereas the out-of-plane one is $J_{\perp} \approx 9-13$ meV for the coupling between the closely spaced $CuO_2$ layers and $J^{'}_{\perp} \approx 0.02-0.4$ meV(14)(19) between the $CuO_2$ bilayers units (across the CuO chains). The spin-waves in the high-Tc cuprates are therefore quasi-two-dimensional and disperse over an energy range of about 300 meV(16).

More recently, resonant inelastic X-ray scattering (RIXS) has emerged as another powerful technique to study the spin wave excitations of the high-Tc cuprates(20). The much larger interaction cross-section of this photon-based technique enables RIXS studies of the magnetic excitations on small single crystals and even on thin films. The RIXS technique also provides the unique possibility to probe the spin excitations (magnons) in an element-specific manner, which is especially helpful for the study of materials which contain different magnetic ions. The reported dispersion of the magnons is quite consistent with that previously reported from INS experiments. In particular, for underdoped $YBa_2Cu_3O_{6.6}$ and $YBa_2Cu_4O_8$, and even for optimally doped and slightly overdoped $Nd_{1.2}Ba_{1.8}Cu_3O_7$ and $YBa_2Cu_3O_7$ cuprates, the RIXS studies have confirmed the persistence of paramagnon excitations due to slowly fluctuating and short-ranged AF spin correlations (21). Notably, a RIXS study of the thickness dependence of the magnon excitations in near optimal doped $NdBa_2Cu_3O_7$ thin films confirmed that the generic behavior of the magnons is similar to that in bulk samples. Remarkably, this holds even for a single unit cell film which was found to exhibit only a moderate reduction of the exchange interaction to $J_{\parallel} \approx 98$ meV (from a bulk-like value of 114 meV of the thicker films)(22).

In the following, we report a corresponding RIXS study of the magnetic excitations in a cuprate/manganite superlattice with ten repetitions of 10 nm $Nd_{0.65}(Ca_{0.7}Sr_{0.3})_{0.35}MnO_3$ (NCSMO) and 7 nm $YBa_2Cu_3O_7$ (YBCO) and a final 10 nm NCSMO cap layer. Bulk $YBa_2Cu_3O_7$ is nearly optimal doped high-Tc superconductor with $T_c \approx 90$K and NCSMO an insulator with a CE-type AF and charge orbital order (COO) that coexists and competes with a ferromagnetic order that is strengthened and prevails in large external magnetic fields(23). The resistance curves of the superlattice (NY-SL) show an onset of the SC transition around $T_c \approx 90$K and the magnetization data reveal a weak ferromagnetic signal below about 120K (that is enhanced by a large magnetic field)(24).

Previous studies with X-ray absorption spectroscopy (XAS) on similar cuprate/manganite heterostructures with ferromagnetic $La_{2/3}Ca_{1/3}MnO_3$ (LCMO) layers have demonstrated that their electronic and magnetic properties in the vicinity of the interfaces are strongly modified. X-ray linear dichroism (XLD) measurements established that the Cu–d electrons of the interfacial $CuO_2$ layers undergo a so-called orbital reconstruction whereby about 50% of the holes are redistributed from the $d_{x^2\text{-}y^2}$ to the $d_{3z^2\text{-}r^2}$ orbitals(25–31). They also revealed a transfer of electrons from the manganite to the cuprate side of the interface that reduces the hole doping of the interfacial $CuO_2$ layer which thus is expected to have a weakened superconducting order and host a static AF order. Moreover, XMCD studies have identified an induced ferromagnetic Cu moment that is antiparallel to the Mn moment(32–34). The above-described phenomena have been phenomenologically described in terms of the hybridization between the Cu-$3d_{3z^2\text{-}r^2}$ orbitals of the Cu and Mn ions via the Cu-$O^{ap}$-Mn bonds ($O^{ap}$: apical Oxygen) across the interface(25).

Previous RIXS studies of such cuprate/manganite multilayers have demonstrated that the Cu-charge density wave (Cu–CDW) order and also the crystal field excitations (dd–excitation) of the interfacial $CuO_2$ layers can be tuned via the hole doping (x) and the tolerance factor of the manganite layers(30). Moreover, a new kind of Cu–based charge order with a large wave length of about ten YBCO unit cells, a sizeable coherence length of 40 nm, and a $d_{z^2}$ character rather than the usual $d_{x^2\text{-}y^2}$ one, has been observed in some of these multilayers(31).

Corresponding RIXS-studies of the magnetic excitations in heterostructures from transition metal oxides are rare and do not have the energy resolution required for a fine-structure analysis as to distinguish between the contributions from the interface and from the central part of the layers(35). In particular, there exists, to our best knowledge, no corresponding RIXS study of the spin order at the cuprate/manganite interfaces. Moreover, little is known about a possibly related orbital order which may accompany the orbital reconstruction that yields similar amounts of holes with $d_{x^2\text{-}y^2}$ and $d_{3z^2\text{-}r^2}$ orbital character. Notably, such a combined AF and orbital order might give rise to a so-called altermagnetic state (36) that has recently obtained great attention since it yields a spin-splitting of the Fermi-surface with nodes and sign changes (37) that enables new device concepts in spintronic and magnonics and may also give rise to novel superconducting proximity effects.

## 1. Results and Discussion

**A. XAS study.** Fig. 1 shows representative X–ray absorption spectroscopy (XAS) spectra at the Cu-$L_3$ edge which confirm that our NY-SL exhibits the same kind of interfacial charge transfer and orbital reconstruction effects as those that were previously reported for corresponding manganite/cuprate heterostructures.(25–29, 31–33). Fig. 1A shows a sketch of the X–ray absorption spectroscopy (XAS) experiment with the X–rays in $\pi$- or $\sigma$- polarization incident at 30° with respect to the film plane. As indicated, the X–ray absorption spectra have been recorded simultaneously in fluorescence yield (FY) and total electron yield (TEY) modes at $4K$. Due to the small escape depth of the excited electrons, the TEY response is governed by the topmost cuprate layer, which, for our NY-SL is the interfacial $CuO_2$ layer located underneath the final NCSMO layer. The FY signal is hardly depth sensitive, on the scale of the layer thickness of tens of nanometers, and thus represents the average response of the Cu ions throughout the SL. Fig. 1B confirms that the latter FY spectra are similar to those reported for bulk YBCO (29, 38, 39). In particular, they exhibit a resonance peak at 931 eV, that is much stronger in ab-polarization than in c-polarization and thus is characteristic of Cu–$3d^9$ holes that reside predominantly in the $d_{x^2\text{-}y^2}$ orbital. The corresponding TEY spectra in Fig. 1C reveal a remarkably different behavior that is characteristic of an interfacial charge transfer and an orbital reconstruction of the interfacial Cu ions. The charge transfer is evident from a substantial red-shift of the resonance peak of the interfacial Cu ions from 931 eV to about 930.4 eV. In section B we will use this characteristic difference in the resonance energies of the bulk-like and the interfacial Cu ions to distinguish their respective magnon excitations as seen in the RIXS experiment. The orbital reconstruction is evident in the XAS spectra from a strong enhancement of the c-axis component of the 930.4 eV peak, for which the intensity is comparable to that of the ab-plane component, signaling that the holes in the interfacial $CuO_2$ layer are more or less equally distributed between the $d_{x^2-y^2}$ and $d_{3z^2-r^2}$ orbital orbitals.

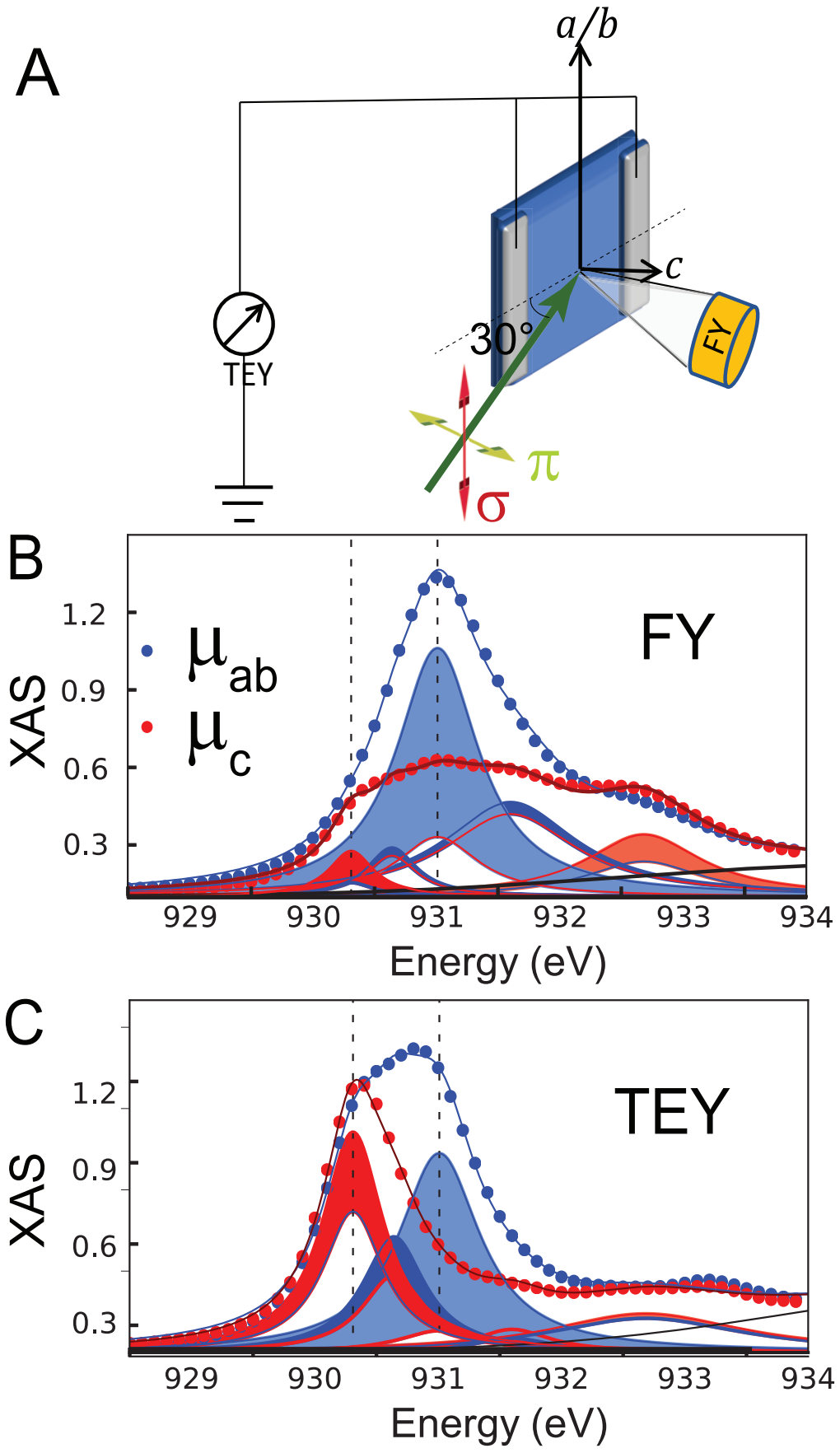

**Fig. 1. X-ray absorption spectroscopy (XAS) data showing the charge transfer and orbital reconstruction effects of the interfacial $CuO_2$ layer.** (A) Schematics of the XAS experiment in fluorescence yield (FY) and total electron yield (TEY) modes. (B) and (C) FY and TEY spectra and multi-peak fits for the in-plane and out-of-plane components of the linearly polarized incident X-rays. Blue (red) shaded areas denote contributions with ($\mu_{ab}$-$\mu_c$)> 0 (< 0), where $\mu_{ab}$ ($\mu_c$) is the net absorption along the in-plane (out-of- plane) direction, as explained in the section S1.1. of the Supporting Information(40).

**B. RIXS Studies.** Next, we turn to the high-resolution RIXS study at the Cu-$L_3$ edge of the NY-SL that has been conducted at the I21 beamline of the Diamond light source at 20K using a grazing exit geometry with $\pi$-polarization of the incident X-ray beam and a scattering angle of 50°, as sketched in Fig. 2A. The RIXS spectra have been measured with an energy resolution of 42 meV. They have been corrected for self-absorption effects and subsequently normalized to the area of the dd–excitations above 1 eV (as shown in Supporting Information Fig. S4 and Fig. S5 (40)). Fig. 2B shows a sketch of the lattice structure of YBCO and NCSMO and of the magnetic structure of the Cu- and Mn- spins in the vicinity of the cuprate/manganite interface. The related structural (green dots) and magnetic (red dots) reciprocal lattice vectors of YBCO are displayed in Fig. 2C, where the dotted lines show the corresponding first Brillouin zones (BZ). Series of RIXS measurements have been performed by varying the sample rotation angle from $\alpha$ =9.8° to $\alpha$ =54.7° (Fig. 2A) as to map out the dispersion along the in-plane momentum directions $[h, 0]$ and $[h, h]$ with respect to the crystallographic BZ of YBCO. Here we use the convention that the vectors $[h, 0]$ and $[h, h]$ are written in units of $a^*$ and $A^*$, respectively, as detailed in Fig. 2C. In agreement with previous RIXS studies (2, 21, 22, 41–47), we assume that the dispersion of the relevant orders and excitations along the out-of-plane direction is very weak and can therefore be neglected.

The crystal field (dd) excitations at higher energy agree rather well with those previously reported for bulk YBCO (44, 48) (see also Supporting Information(40) section S2.2. and Fig. S6). In the following, we focus on the low-energy region of the RIXS spectra (below −0.6 eV) which contains, besides the elastic peak, contributions from inelastic excitations, like phonons, magnons and bimagnons. More details about the fitting of these low-energy features can be found in section S2.2 of the Supporting Information(40).

Fig. 2D shows a representative RIXS spectrum below −0.6eV at the bulk Cu resonance energy of 931 eV and a momentum vector of $0.3A^*$. It contains two high energy phonons around 60 and 80 meV (gray shading) that can be assigned to the buckling and breathing modes of YBCO, respectively(49). The phonon modes at lower energy are not resolved and thus contribute to the elastic peak (gray line).Following the analysis of a previous high-resolution RIXS study of $NaBa_2Cu_3O_6$ (50), the fairly weak and broad peak around 350 meV (dark green line) is assigned to a bi-magnon excitation. Notably, the spectrum contains two strong, additional modes with maxima around 120 and 200 meV (blue and red shadings) which dominate the signal above the phonon range at $\Delta E > 80$ meV. In the following we show that both modes are of magnetic origin, and, accordingly, we denote them as M1 and M2 modes. Moreover, we provide evidence that the M2 mode corresponds to the magnon of the bulk-like $CuO_2$ layers that are located away from the interface whereas the M1 mode arises from the magnetic excitations of the interfacial $CuO_2$ layers. This assignment is supported by the comparison of Fig. 2D and 2E, which show the RIXS spectra at the momentum transfer $0.3A^*$ for the incident photon energies of 931 eV and 930.5 eV at the resonances of the bulk-like and the interfacial Cu ions, respectively (see the XAS spectra in Fig. 1). This comparison highlights that the M2 mode is the most pronounced feature in the spectrum at 931 eV (bulk resonance) whereas the M1 mode prevails at 930.5 eV (interfacial resonance).

To ascertain the magnetic nature of the M1 and M2 modes, we performed additional RIXS-polarimetry measurements at the ID32 beamline of ESRF (51) at 20K. This experiment resolves the polarization of the incident and of the scattered X-rays. Accordingly, it allows one to distinguish between the non-spin flip ($\pi$-$\pi$) and the spin-flip ($\pi$-$\sigma$) signals of which only the latter arises from scattering with a single magnon. As shown schematically in Fig. 2A, the polarization analysis is done by inserting after the analyzer a mirror with different reflection coefficients for $\sigma$- and $\pi$-polarized X-rays. Fig. 2F shows the obtained polarization resolved spectra at $0.3A^*$, after self–absorption correction and normalization. It confirms that the signal in the energy–loss range of the M1 and M2 peaks is governed by the spin–flip channel and is therefore predominantly due to single-magnon scattering. Except for the low-energy range with the elastic peak and the phonons (below about 100 meV), the non-spin-flip ($\pi$-$\pi$) signal in Fig. 2D is much lower than the spin-flip one. Only two weak peaks around 0.38 eV and 0.2 eV, that agree well with the bi-magnons of the fits in Fig. 2D, rise here above the calculated error bar.

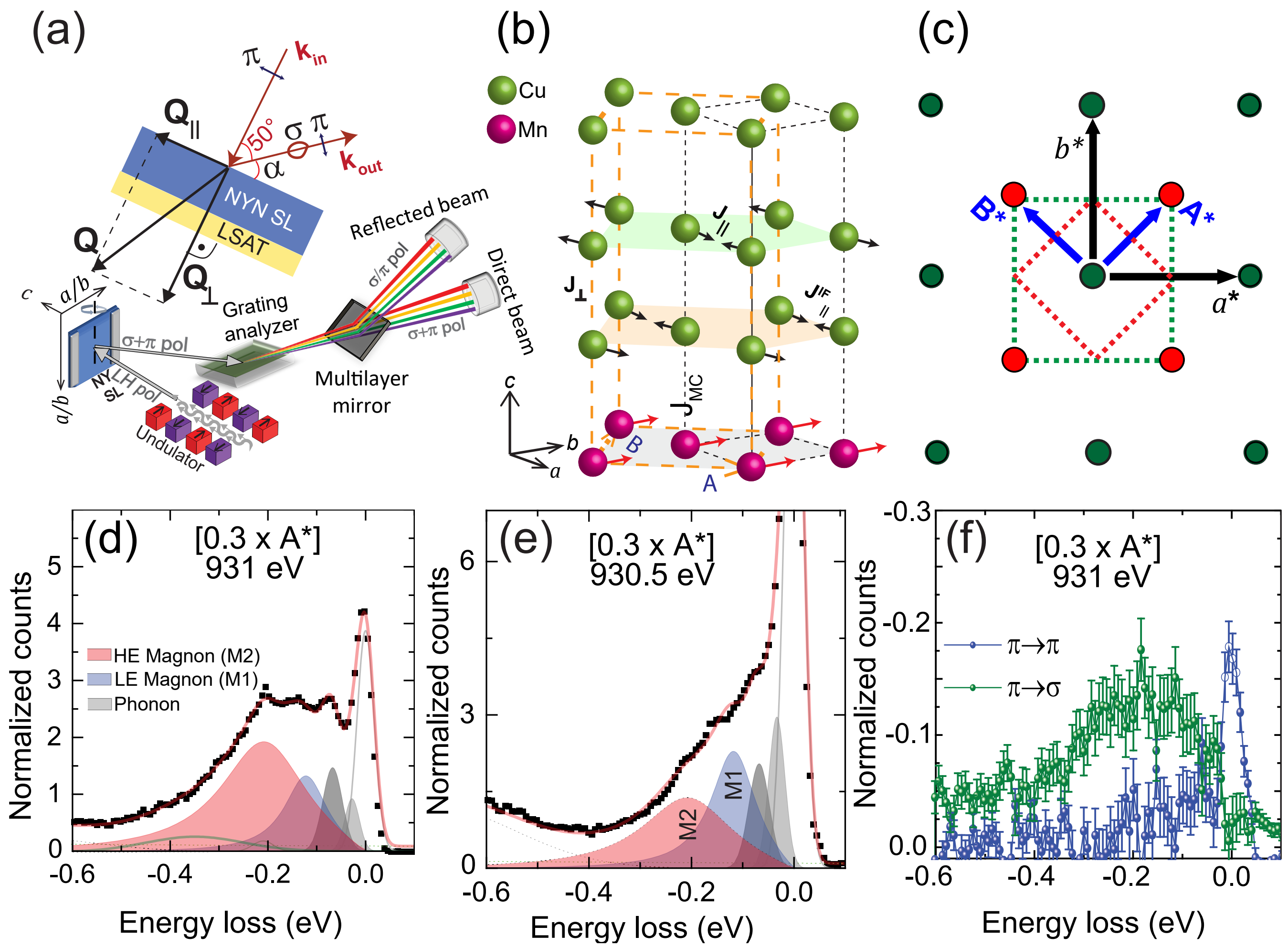

**Fig. 2. RIXS study of the distinct magnon modes of the bulk-like and interfacial $CuO_2$ layers.** (A) Sketch of the RIXS and polarimetry experiments indicating the measurement geometry. (B) Scheme of the magnetic order and exchange interaction of the Cu- and Mn-spins in the vicinity of the cuprate/manganite interface. (C) Reciprocal lattice vectors of the structural (green dots) and the magnetic (red dots) orders with the corresponding first Brillouin zones (BZ) shown by dotted lines. (D) and (E) Representative RIXS spectra at $[0.3 \times A^*]$ showing the two magnon modes M1 (blue shading) and M2 (red shading) along with phonons (gray shading) and the elastic line (grey line) at the bulk-like Cu resonance of $931$eV and the interfacial Cu resonance of $930.5$ eV, respectively. The spectra are normalized with respect to the area of the dd–excitations beyond $-1$ eV (not shown); (f) Corresponding RIXS-polarimetry spectra at $[0.3 \times A^*]$ and $931$ eV showing that the M1 and M2 modes are predominantly from spin-flip scattering and thus of magnetic origin. The error bars have been calculated as described in section S3.2 of the Supporting Information (40).

With this information at hand, we analyzed the full, angle-dependent series of RIXS spectra (without polarization analysis of the scattered beam) at the bulk-like resonance of 931 eV and near the interface resonance of 930.5 eV. Both series have been fitted simultaneously using the two magnon modes M1 and M2. Since the YBCO layers in our sample are optimally doped, the magnons are expected to be fairly broad and resemble the response function of an overdamped harmonic oscillator(52).

The measured RIXS spectra and the corresponding fits are displayed in Fig. 3, the dispersion of the best fit parameters of the magnon modes M1 (blue hollow squares) and M2 (red solid dots) are summarized in Fig. 4. The parameters of the M2 mode and their dispersion along the $[h, 0]$ and $[h, h]$ directions are similar to those reported for bulk RBCO and corresponding thin films(2, 21, 43, 45). Specifically, the energy of the M2 mode has a maximal value of about 300 meV at the largest wave vector of $0.43a^*$ and it decreases continuously towards smaller values of $h$. Moreover, the intensity of the M2 mode shows a strong decrease towards small $h$ values similar to the bulk magnons in AF YBCO (2, 21, 43, 45, 49). To the contrary, the M1 mode has a much lower maximal energy of about 120 meV and is only weakly dispersive. Notably, the evolution of the intensity of the M1 mode is opposite to that of the M2 mode, i.e. it is weakest at large $h$ values and increases strongly as $h$ decreases. The distinct resonance energy, the weakly dispersive behavior, and the unusual spectral weight increase towards small in-plane momentum transfer thus clearly distinguish the M1 mode from the usual magnon excitations in YBCO or related planar high-Tc cuprates(21, 49, 53).

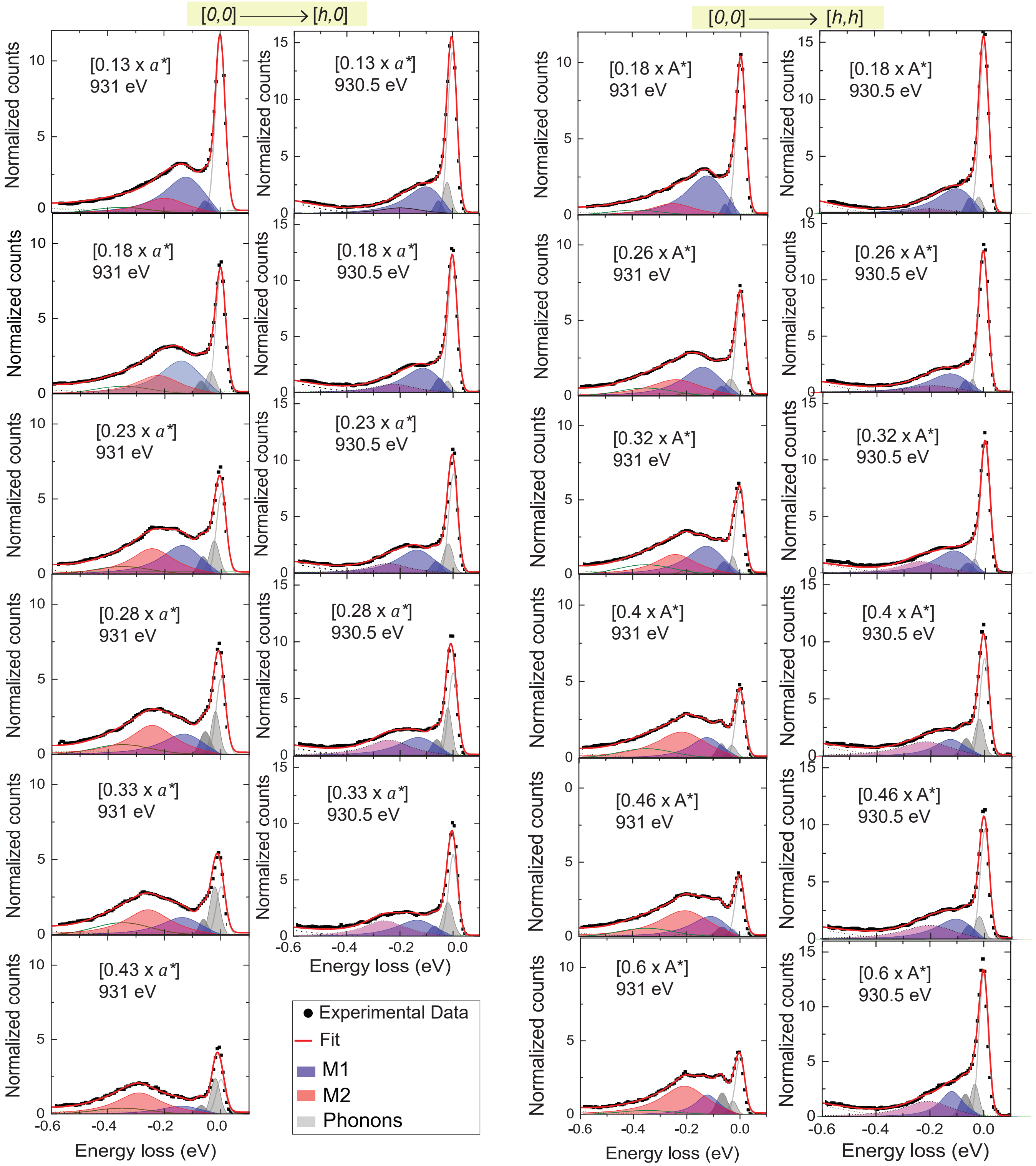

**Fig. 3.** Comparison of the RIXS spectra (black dots) and their fitting (solid lines and color shading) at the bulk resonance energy of $931$ eV and the interface resonance energy of $930.5$ eV along $[h, 0]$ and $[h, h]$. The contributions of the M1 and M2 modes are shown by the blue and red shadings, respectively. The legend is presented at the bottom of the 2$^{nd}$ column.

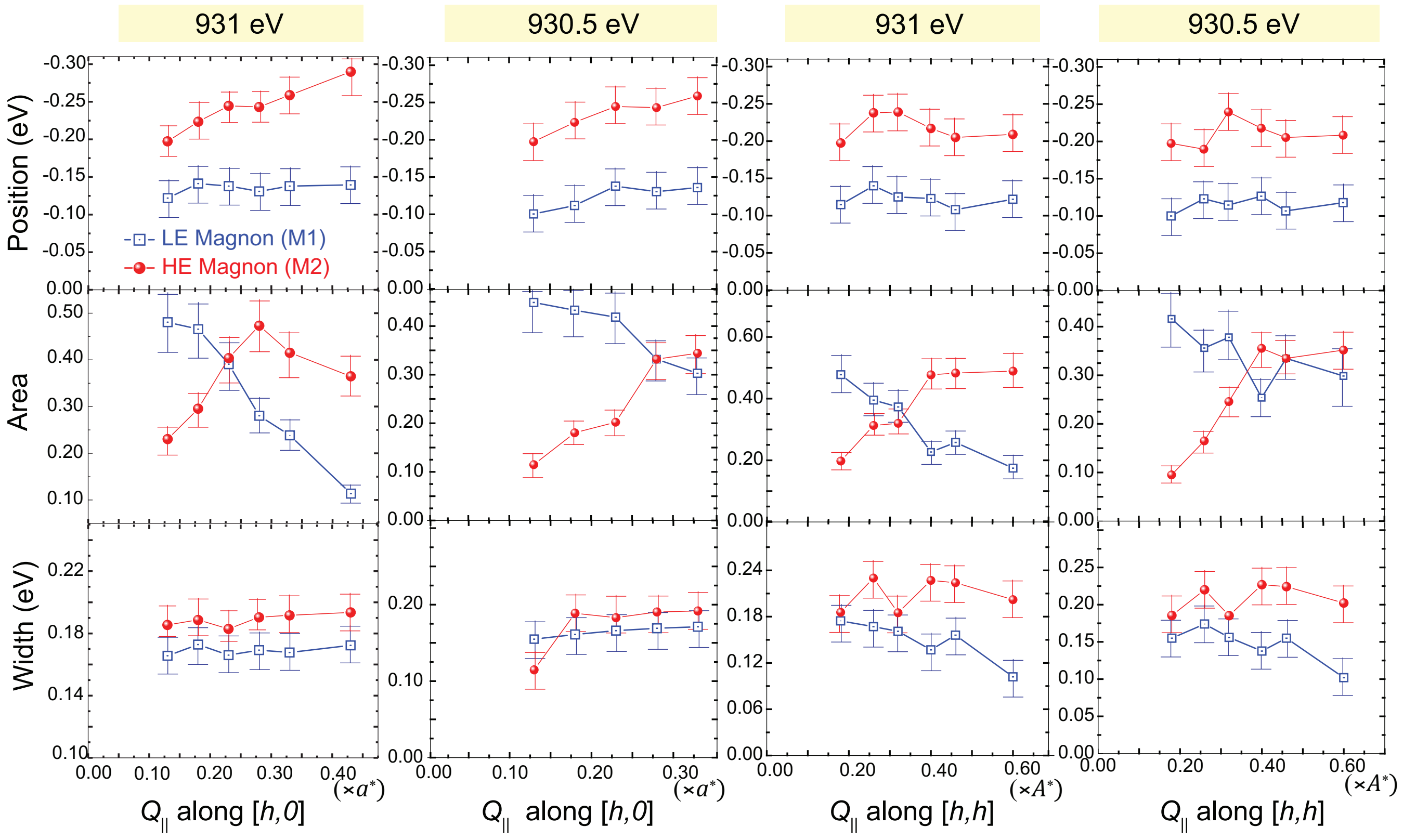

**Fig. 4.** Summary of the best fit parameters obtained for the dispersion of the magnon modes M1 and M2 along $[h, 0]$ and $[h, h]$ at the bulk resonance energy of $931$ eV and the interface resonance energy of $930.5$ eV. Calculation of the error bars is discussed in section S2.3 of the Supporting Information(40).

## 2. Comparison of RIXS data with linear spin wave model

**A. Linear spin wave model.** In the following, we show that a minimal model based on linear AF spin wave theory can account for the main features of the M1 and M2 modes, except for the unusual intensity increase of the M1 mode toward small wave vectors. The schematics of this minimal spin wave model is depicted in the Fig. 2B. For the $CuO_2$ layers that are not located right at the cuprate/manganite interface we adopt a 2D Heisenberg nearest neighbor model with an in-plane AF exchange interaction, $J_{\parallel}$ , and an interplanar AF exchange within the $CuO_2$ bilayer units, $J_{\perp} \ll J_{\parallel}$ , similar to that in bulk YBCO. For the interfacial $CuO_2$ bilayer unit we assume that only the $CuO_2$ layer that is directly bonded via an apical oxygen to the adjacent $MnO_2$ layer has strongly modified electronic, orbital and magnetic properties. Because the RIXS spectra can be fitted well with only two magnon modes, we assume that the second $CuO_2$ plane of this bilayer unit has already bulk-like properties.

Finally, our minimal spin wave model also includes a weak AF exchange coupling across the interface with the adjacent $MnO_2$ layer, $J_{MC}$ . However, as detailed in the supporting material in section S1.3 and S4.1, it turns out that $J_{MC}$ is very small, i.e. below 1meV, and thus affects the magnon dispersion of the interfacial $CuO_2$ layer only at very low energies that are not relevant for the analysis of the dispersion of the magnon data (M1 mode) that is presented in the following paragraph.

**B. Fitting of magnon dispersion.** Fig. 5 shows a comparison of the dispersion of the M1 and M2 modes of the RIXS data with that of the best fits with the above-described minimal spin wave model for which the M1 and M2 modes are assigned to the magnons of the interfacial and the bulk-like $CuO_2$ layers, respectively. Figs 5A and 5B compare the dispersion of the magnon energy along the $[h, 0]$ and $[h, h]$ directions, respectively.

For the fitting, only the in-plane exchange parameters ($J_{\parallel}$ and $J^{IF}{}_{\parallel}$ ) were allowed to vary. To keep the number of fitting parameters at a minimum, we fixed the out-of-plane interaction of the bulk-like $CuO_2$ layer to $J_{\perp} = 7$ meV (as in bulk YBCO). Likewise, we used a fixed value of $J_{MC} = 0.5$ meV for the exchange coupling between the interfacial $CuO_2$ and $MnO_2$ layers. Note that both of these small out-of-plane exchange parameters do not have a noticeable effect on the magnon dispersion curves in the relevant energy range of the M1 and M2 modes above 100 meV.

The best fit of the in-plane exchange interactions yields values of $J_{\parallel} =$ 130 meV and $J^{IF}{}_{\parallel} \approx$ 70 meV. The former parameter, that determines the dispersion of the M2 mode in the bulk-like $CuO_2$ layers, is indeed similar to that reported for bulk $[Re]Ba_2Cu_3O_7$ or thin films of $[Re]Ba_2Cu_3O_7$(13, 16, 21, 22, 41). The latter parameter signals a strong suppression of the in-plane AF exchange coupling of the interfacial $CuO_2$ layer. As discussed in Ref.(28), such a large suppression of $J^{IF}{}_{\parallel}$ is expected from the orbital reconstruction of the interfacial $CuO_2$ layer, which increases the population of holes on the Cu-$3d_{3z^2-r^2}$ orbitals and thus gives rise to $Cu(d_{3z^2-r^2})-O-Cu(d_{x^2-y^2})$ and $Cu(d_{3z^2-r^2})-O-Cu(d_{3z^2-r^2})$ bonds that have a lower hopping probability than the $Cu(d_{x^2-y^2})-O-Cu(d_{x^2-y^2})$ bonds that prevail in the bulk-like $CuO_2$ layers.

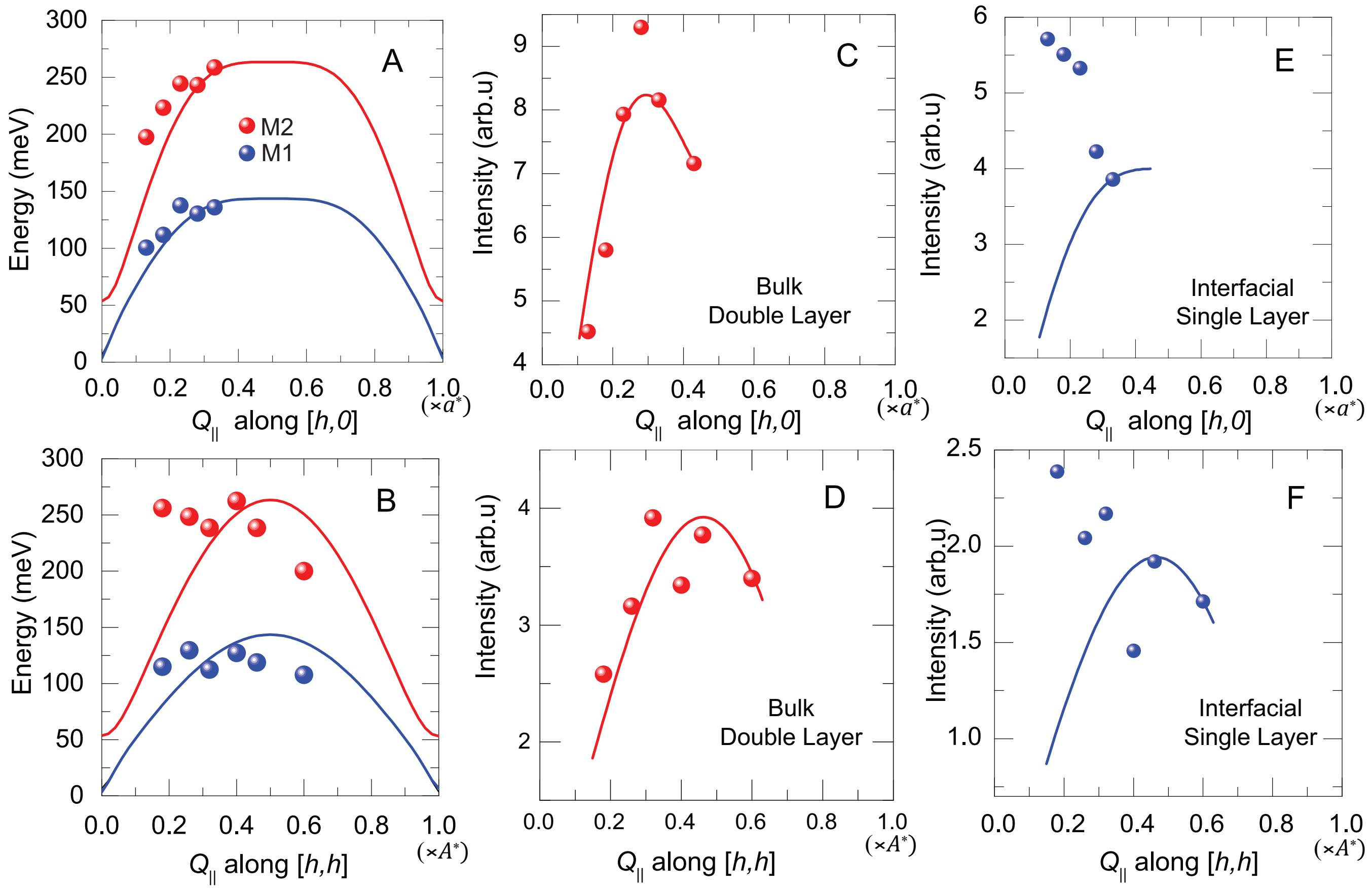

**Fig. 5. Fitting of the magnon dispersion in the bulk-like and interfacial $CuO_2$ layers with a linear spin-wave model.** (A) and (B): Dispersion of the magnon energy along the ($\pi$,0) and ($\pi$,$\pi$) directions. (C) and (D) Corresponding dispersion of the intensity of the bulk-like magnon and (E) and (F) of the magnon of the interfacial $CuO_2$ layer. Note that the calculated intensities have been scaled at the highest wave vector to those of the experimental data point.

Figs 5C,D and 5E,F display the comparison of the measured (symbols) and calculated (lines) dispersion of the intensities of the magnon modes for the bulk-like and the interfacial $CuO_2$ layers, respectively. Since the absolute values of the experimental intensities cannot be quantified, the comparison with the theoretical values has been facilitated by rescaling them so they are matched at the largest wave vector of the measurement. For the M2 mode from the bulk-like $CuO_2$ layers this yields a fairly good agreement between the experimental and the calculated magnon intensities. To the contrary, for the M1 mode there exists a striking discrepancy between the measured and the calculated magnon intensities. Whereas the linear spin wave model predicts that the magnon intensity should decrease and finally vanish towards $[0,0]$, the experimental magnon intensity exhibits a steep increase towards small wave vectors. This striking discrepancy reveals that our minimal spin wave model is lacking an important feature of the magnetic and electronic state of the interfacial $CuO_2$ layer. In the next paragraph, we show that the missing feature turns out to be an additional electronic order that is coupled in a very specific way with the antiferromagnetic order of the interfacial $CuO_2$ layer.

**C. Combined AF and orbital order of the interfacial $CuO_2$ layer.** In the following, we show that the above described discrepancy between the measured evolution of the intensity of M1 mode and that predicted by the minimal spin wave model can be readily resolved in terms of an additional spatial order that develops concurrently with the AF order in the interfacial $CuO_2$ layer. To account for the intensity increase of the M1 mode towards small wave vectors, this additional order needs to give rise to a strong modification of the amplitudes of the RIXS matrix elements for the spin-up and spin-down components of the AF state. As outlined in section S4.3 and Eq. S4.16 and Eq. S4.21 of the Supporting Information(40), this allows to remove the destructive interference effect on the RIXS intensity that occurs in the plain AF state for which the spin-up and spin-down components develop a 180°-phase shift at small scattering wave vectors.

A natural candidate for causing the required modification of the scattering amplitudes of the spin-up and spin-down components of the AF state is a lateral order of the $d_{x^2-y^2}$ and $d_{3z^2-r^2}$ orbitals along the interface that correlates specific spin and orbital states. Fig. 6A displays a sketch of the simplest and most likely case of a combined AF and checkerboard-type orbital order for which the spin-up states are correlated with the $d_{x^2-y^2}$ orbitals and the down spins with the $d_{3z^2-r^2}$ orbitals (or vice versa). Fig. 6B shows the accordingly calculated momentum dependence of the intensity of the magnon mode (solid line) for which the individual scattering amplitudes of the $d_{x^2-y^2}$ and $d_{3z^2-r^2}$ orbitals have been derived from a single ion model, as described in section S4.3 of the Supporting Information. Notably, this model of a combined AF and checkerboard orbital order reproduces very well the anomalous increase of the intensity of the M1 mode towards small wave vectors. Note that within this simple spin-wave model, the additional orbital order merely affects the scattering amplitudes of the spin-up and spin-down components of the AF order of the interfacial layer. The corresponding effect on the exchange interaction of the interfacial $CuO_2$ layers, and on the resulting dispersion of the M1 mode, has been already accounted for in terms of the reduced value of $J^{IF}{}_{||}$ , as discussed in the previous section. Also note that the energy and intensity dispersion of the M2 mode from the bulk-like $CuO_2$ layers is hardly affected by the interfacial orbital order, due to the very weak interlayer exchange interaction and the resulting quasi-two-dimensional spin wave properties. Accordingly, in the relevant high energy range above about 50 meV, the energy dispersion of the M1 and M2 modes as shown in Figs. 5A and 5B, as well as the intensity variation of the M2 mode in Figs. 5C and 5D, are not noticeably affected by this interfacial orbital order.

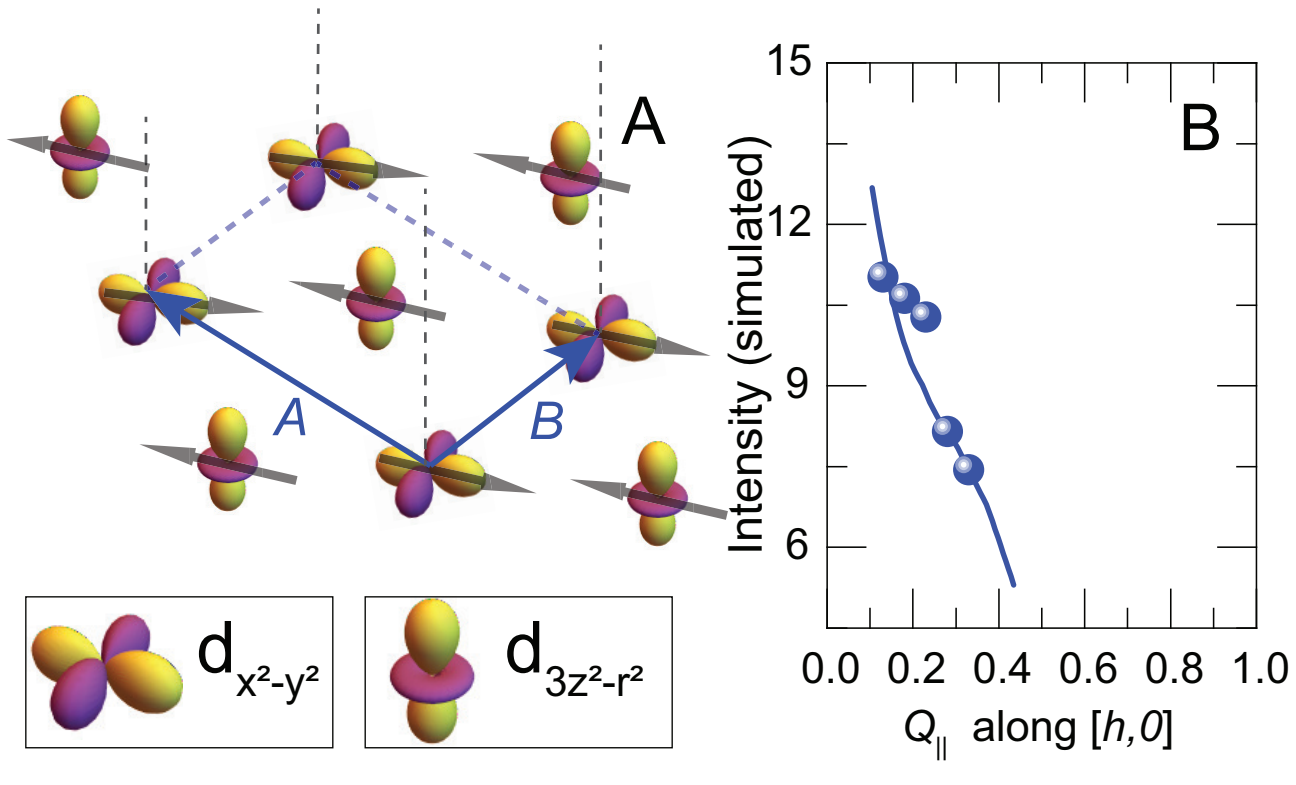

**Fig. 6. Evidence for a combined magnetic and orbital order at the interfacial $CuO_2$ layer.** (A) Schematic of the proposed combined AF and checkerboard orbital order at the interfacial $CuO_2$ layer. (B) Comparison of the dispersion along $[h, 0]$ of the intensity of the M1 mode (dots) with that calculated for the model shown in panel (A).

Naturally, the required disparity of the scattering amplitudes for the spin-up and spin-down components of the AF order could also be caused by a more complex orbital order or even by another type of order, such as a charge order which modulates the density of holes on the spin-up and spin-down sites. Nevertheless, we consider the orbital order described in Fig. 6A to be the most likely candidate, especially since it conforms with the more or less equal number of holes on the $d_{3z^2-r^2}$ and $d_{x^2-y^2}$ orbitals of the interfacial $CuO_2$ layer that is suggested by the TEY-XAS spectra in Fig. 1C.

Irrespective of the detailed nature of the additional electronic order of the interfacial $CuO_2$ layer, it turns out that all the relevant combined AF and electronic (charge or orbital) orders should give rise to a so-called altermagnetic state (54). Such altermangets are compensated, collinear magnets which exhibit a characteristic spin-splitting of the Fermi-surface with nodes at which the polarization changes its sign, such that the average spin-polarization around the Fermi-surface vanishes. Notably, this spin splitting can be sizable irrespective of the spin-orbit-coupling strength. The combined AF and checkerboard orbital order shown in Fig. 6A corresponds to a so called *d*-wave magnet for which the spin-splitting of the Fermi-surface exhibits a d-wave symmetry. Here it is evident from symmetry considerations that the difference in energy of the $d_{3z^2-r^2}$ and $d_{x^2-y^2}$ states (in an orthorhombic or tetragonal environment) and of the respective wave functions with spin up and spin down results in an anisotropic spin polarization in momentum space.

## 3. Summary and conclusions

In summary, we performed a high-resolution and polarization-resolved RIXS study at the Cu-$L_3$ edge to investigate the magnetic and orbital orders at the $YBa_2Cu_3O_7/Nd_{0.65}(Ca,Sr)_{0.35}MnO_3$ interface. We found that the RIXS spectra contain two distinct magnon modes, denoted as M1 and M2, that can be assigned to the $CuO_2$ layers right at the interface (M1 mode) and those located further away from it (M2 mode), according to their characteristic resonance energies at 930.5 and 931 eV, respectively. By comparing the measured magnon dispersion with calculations based on a minimal linear spin wave model, we confirmed that the M2 mode is well described with a bulk-like parameter of the in-plane exchange interaction $J_{||} \approx 130$ meV. The corresponding fit of the energy dispersion of the M1 mode yields a strongly reduced value of the in-plane exchange coupling of the interfacial $CuO_2$ layer of $J^{IF}{}_{||} \approx 70$ meV. We have outlined that such a strong suppression can be readily understood in terms of the orbital reconstruction which, according to the XAS data, yields similar amounts of holes with $d_{x^2-y^2}$ and $d_{z^2}$ orbital character. This results in a high density of $Cu(d_{3z^2-r^2})-O-Cu(d_{x^2-y^2})$ bonds for which the hopping probability (and thus the exchange interaction) is strongly reduced as compared to the $Cu(d_{x^2-y^2})-O-Cu(d_{x^2-y^2})$ bonds of bulk-like YBCO. Nevertheless, a major discrepancy with respect to the prediction of our minimal spin wave model has been observed for the momentum dependence of the intensity of the M1 mode, which has been found to strongly increase towards small momentum transfer even though a strong suppression is predicted by the model. We have outlined that this contradiction can be readily resolved in terms of an additional electronic or orbital order that accompanies the AF order. This additional order needs to be of a specific kind such that it yields different amplitudes of the spin-up and spin-down components of the RIXS matrix elements as to overcome the destructive interference from their 180°-phase shift towards small wave vectors that occurs for the plain AF order. As a likely candidate, that naturally accounts for the unusual momentum dependence of the M1 mode intensity, we have identified a combined AF and checkerboard order of the $d_{x^2-y^2}$ and $d_{3z^2-r^2}$ holes. This checkerboard orbital order conforms with the XAS data which suggest that the $d_{3z^2-r^2}$ and $d_{x^2-y^2}$ orbital character of the holes in the interfacial $CuO_2$ layer is almost equally balanced. Nevertheless, there exist other types of orbital orders, or orders that involve the charge rather than the orbital degree of freedom, that can give rise to a strong disparity of the scattering amplitudes of the up and down spins of the AF state and thus can explain the anomalous intensity evolution of the M1 mode. However, it turns out that all of the relevant cases give rise to a combined AF and charge or orbital order which has the properties of a so-called altermagnetic state(54).

Such altermagnetic orders have recently obtained a great deal of attention since they exhibit a characteristic spin-splitting of the Fermi-surface with nodes at which the sign of the polarization changes. For the specifically discussed AF and checkerboard orbital order shown in Fig. 6A, the spin-splitting of the Fermi-surface varies according to a d-wave symmetry pattern such that it can be classified as a quasi-two-dimensional d-wave magnet. The electronic and magnetic properties of such altermagnets are a fairly new research field that is rapidly growing since they offer unique opportunities for new kinds of spintron and magneto-calorimetric applications(54–56). For example, the directional dependence of their spin-polarization effects in transport and tunneling has been predicted to enable new device functionalities. It has also been pointed out that the interplay between an altermagnet and a superconductor can give rise to unconventional proximity effects or even to a new kind of superconducting pairing mechanism(57, 58). Our observation that the $CuO_2$ layers at a cuprate/manganite interface seem to host a quasi-two-dimensional altermagnetic state, that is in direct proximity to the high-Tc superconducting order in the neighboring $CuO_2$ layer, is therefore of great current interest. In the first place, it calls for the fabrication of suitable devices that enable transport studies of the directional anisotropy of the superconducting currents and/or of the spin polarisation of the normal currents, in order to confirm the altermagnetic properties of this cuprate/manganite interface.

Hopefully, our results will also simulate further studies of the origin of the orbital ordering of the interfacial Cu ions. At present, we can only speculate that it might be driven by a buckling of the interfacial Ba−O layer which helps to accommodate the lattice mismatch between the cuprate and manganite layers. Alternatively, it may be triggered by a corresponding charge/orbital order on the manganite side of the interface. The latter scenario could be further explored with complementary RIXS studies at the Mn-$L_3$ edge. Of great interest would also be corresponding studies of YBCO/manganite heterostructures for which the manganite layers have a higher doping level of $x = 0.5$ and host a well-developed combined CE-type AF and charge/orbital order. Notably, for such heterostructures the XAS experiments have indicated weaker charge transfer and orbital reconstruction effects(24). Moreover, a previous RIXS study has revealed a charge order with a period of about ten unit cells that appears to have a $d_{z^2}$ orbital character(24) which clearly distinguishes it from the charge density wave of bulk YBCO that involves primarily the $d_{x^2-y^2}$ states.

Last but not least, our results have highlighted the potential of the high resolution RIXS technique to selectively probe the magnetic properties at the interfaces of other kinds of magnetic multilayers. Potential candidates are heterostructures from complex oxides that are known to exhibit various kinds of magnetic orders(34, 59, 60), or from other materials with versatile electronic and magnetic properties that are strongly modified in the vicinity of their interfaces, such as the transition metal chalcogenides or related van der Waals materials(61, 62).

## 4. Experimental Methods

**A. Sample preparation and Characterization.** The sample was deposited by Pulsed Laser Deposition. A detailed description of the growth conditions and its characterization is presented in ref(24)

**B. X-ray Absorption Spectroscopy.** The X-ray Absorption Spectroscopy (XAS), X-ray Linear and Magnetic Circular Dichroism (XLD/XMCD) (Fig. 1 and Fig S2 and S3 in Supporting Information) were measured at the XMCD end-station of the ID 32 beamline at ESRF, Grenoble in France(63), in Total Fluorescence Yield (FY) mode and Total Electron Yield (TEY mode), at the Cu-$L_3$-edge at 20K. The TFY detector is a photodiode shielded with an Al foil, in order to avoid electrons reaching the detector. The incidence angle was fixed to 30° with respect to the incoming beam, which was horizontally or vertically polarized. The resolution of the acquisition was 110 meV. Multiple scans were averaged to produce the final plots in Fig. 1, Fig. S2 and S3. The horizontal spot size of the beam was approx. 30 $\mu$m.

**C. Resonant Inelastic X-ray Scattering at I21 beamline in DLS, Oxford.** Resonant Inelastic X-ray Scattering (RIXS) experiments at the Cu-$L_3$ edge with a very high energy resolution of $\sim$ 42 meV have been performed using the RIXS spectrometer at the I21 beamline of the DLS, in Oxford, UK. The measurements were performed in grazing exit geometry at 20K. For every scan, the position of the elastic line, the energy to pixel ratio and the resolution have been determined from a reference measurement on a carbon tape (which acts as a non-resonant scatterer). The obtained energy calibration was 6.7 meV/pixel. The size of the beam-spot was about 2 $\mu$m in the vertical and 40 $\mu$m in the horizontal direction.

**D. RIXS polarimetry at ID32 beamline in ESRF, Grenoble.** The RIXS polarimetry experiments at the Cu-$L_3$ edge have been performed at the ID32 beamline of the ESRF, in Grenoble(51), France. Here the polarization of the scattered beam has been analyzed to determine the magnetic spin-flip scattering. The polarization of the scattered X-ray photons was determined by inserting in the path of the analyzed beam a mirror with a different reflectivity for $\sigma$ and $\pi$ polarization ($R_\sigma$=0.141, $R_\pi$=0.086) and comparing the scattered intensity to that without the mirror. Further details are given in section S3.1 of the Supporting Information(40). For each scan, the position and resolution of the zero-loss position has been tracked using a non-resonant scatterer. The energy calibration for with and without the mirror was 21.471 meV/pixel and 21.306 meV/pixel, respectively. The beam-spot size was 2 $\mu$m in the vertical direction and 40 $\mu$m in the horizontal one.

**Supporting Information Appendix (SI).** This article contains supporting information.

**Data Availability.** The XAS data are available at https://doi.esrf.fr/10.15151/ESRF-ES-511177236.
The RIXS data from ESRF are available at https://doi.esrf.fr/10.15151/ESRF-DC-1314300736 and https://doi.esrf.fr/10.15151/ESRF-ES-518219483.
The RIXS data from DLS are available at https://doi.org/10.5281/zenodo.8283049.

**ACKNOWLEDGMENTS.** We acknowledge helpful discussions with Bruce Norman, Dominik Munzar and Jiri Chaloupka. Christian Bernhard and Subhrangsu Sarkar: Swiss National Science Foundation (SNSF) through Grant No. 200020_172611. C.M. and C.W.N. acknowledge financial support from the Swiss National Science Foundation (SNSF) Grant No. P00P2_170597. Yu. G. Pashkevich acknowledges the financial support of the Swiss National Science Foundation through the individual grant IZSEZ0_212006 of the ≪Scholar sat Risk≫ Program. R.C acknowledges the support by a grant of the Ministry of Research, Innovation and Digitization, CNCS-UEFISCDI, project number PN-III-P1-1.1-PD-2021-0238, within PNCDI III and a scholarship "Cercetare postdoctorală avansată" funded by the West University of Timisoara, Romania. Last but not least, the authors acknowledge the generous help and excellent infrastructure provided by DLS and ESRF.

## Supporting Information for

# Composite antiferromagnetic and orbital order with altermagnetic properties at a cuprate/manganite interface

Subhrangsu Sarkar, Roxana Capu, Yurii G. Pashkevich, Jonas Knobel, Marli Dos Reis Cantarino, Abhishek Nag, Kurt Kummer, Davide Betto, Roberto Sant, Christopher W. Nicholson, Jarji Khmaladze, Ke-Jin Zhou, Nicholas B. Brookes, Claude Monney, Christian Bernhard

Paste corresponding author name here
Email: subhrangsu.sarkar@unifr.ch, roxana.capu@e-uvt.ro, christian.bernhard@unifr.ch

**This PDF file includes:**

- Supporting text
- Figures S1 to S8
- SI References

# Supporting Information Text

## S1. X-ray absorption Spectroscopy

### *S1.1. Linear dichroism and multi-peak fitting:*

For each polarization of the incident X-ray beam, the experimental spectra are first averaged and a linear background is subtracted such that the intensity well below the Cu-$L_3$ edge vanishes. Subsequently, the intensity is normalized to yield a unity value for the edge jump above the Cu-$L_3$ edge. The spectrum for linear $\sigma$ polarization of the incident X-ray beam represents the response along the ab-plane of YBCO, $\mu_{ab}$. The corresponding c-axis response, $\mu_c$, has been derived from the measured spectrum in linear $\pi$-polarization, $\mu_\pi$, and the incidence angle of the x-ray beam of 30° according to the following equation(1):

$$\mu_c = 1/\,(\cos^2(30°)\mu_\pi - \tan^2(30°)\mu_{ab})$$

The corresponding polarization-averaged absorption, $\mu_{avg}$, amounts to:

$$\mu_{avg} = \max\frac{(2\mu_{ab} + \mu_c)}{3}$$

In Fig. 1 in the main manuscript, the solid symbols represent the experimental data and solid lines show the best fits with Lorentzian functions. The colored shading indicates when $\mu_{ab} > \mu_c$ ($\mu_{ab} < \mu_c$) and the difference ($\mu_{ab} - \mu_c$) is shaded in blue (red) indicating a positive (or negative) value of the x-ray linear dichroism. During the fits, we fixed the position of the peaks in the FY and TEY spectra with a maximal deviation of 0.1eV and fitted both of them concomitantly.

### *S1.2. XMCD spectra:*

The X-ray magnetic circular dichroism (XMCD) signal (also measured at an incidence angle of 30°) is proportional to the difference between the normalized spectra measured with right circular polarization, $\mu_+$, and left circular polarization, $\mu_-$, (measured at same temperature and applied magnetic field) according to the following equation:

$$XMCD = \frac{(\mu_+ - \mu_-)}{\frac{1}{2}(\mu_+ + \mu_-)} \times 100$$

Here, $\frac{1}{2}(\mu_+ + \mu_-)$ corresponds to the XANES spectrum that is shown in the Fig S2 and S3 for the measurements in TEY (Cu and Mn L3-edge) and FY (Cu-L3 edge) modes.

### *S1.3. Experimental observation of hysteretic Cu-moment at the interface:*

The Cu-XMCD data in Fig. S2 reveal a weak FM Cu moment that originates from the interfacial Cu ions. This assignment is based on the distinct resonance energies of the bulk-like and the interfacial Cu ions at 931 eV and 930.5 eV, respectively, which are evident from the XAS spectra in FY and TEY mode shown in Fig. S2A. Fig. S2B reveals that the Cu-XMCD signal is indeed peaked around 930.5 eV and thus at the resonance of the interfacial Cu ions. Figs. S3 confirms that in FY mode this Cu-XCMD signal is also peaked at 930.5 eV and also considerably weaker than in TEY mode.

The FM Cu moment can be understood in terms of a canting of the planar AF order of the interfacial $CuO_2$ layer that is induced by an AF exchange interaction with the FM Mn moments on the manganite side of the interface. The energy minimization in Heisenberg approximation suggests that the induced Cu moment is perpendicular to the Néel vector and lying within the $CuO_2$ plane, as sketched in Fig. S3D.

The AF nature of the exchange interaction and the subsequent antiparallel alignment of the Cu moment with respect to the Mn moment is evident from the opposite signs of the Cu- and Mn-XMCD signals in a small magnetic field of 0.5 Tesla shown in Figs S2B and S2F, respectively. The magnitude of $J_{MC}$ can be estimated from the characteristic magnetic field dependence of the Cu-XMCD signal in Figs S2B and S2C which exhibits a sign reversal at a critical field of $H_c \sim 3-4$ Tesla above which it becomes positive as the Zeeman energy overcomes the AF exchange coupling, $J_{MC}$ , and thus causes a reorientation of the Cu moment. According to our model calculations (see Eq.S4.5 below) this crossing point thus yields an estimate of $J_{MC} \sim \mu_B \times H_c \sim 0.5$ meV.

Such a small value of $J_{MC}$ is also suggested by the analogy with the exchange coupling between Cu spins on the planar and the chain sites of the undoped and antiferromagnetic $YBa_2Cu_3O_6$, which is theoretically predicted around 1 meV(2).

Notably, a corresponding field-induced reorientation of the Cu moment was observed in YBCO/$La_{2/3}Ca_{1/3}MnO_3$ heterostructures in Ref.(3), where it was shown that the strength of the interfacial exchange coupling $J_{MC}$ can vary substantially for samples with different electronic and magnetic properties of the LCMO layers. Moreover, it was previously shown with XRMR that the Cu moments that give rise to the FM order are located on the cuprate side of the interface and thus do not arise from interdiffusion of some Cu ions across the interface during the PLD growth(4). A more detailed understanding of how $J_{MC}$ and the induced FM Cu moment vary with the electronic and magnetic properties of the manganite layers, that can be strongly modified via the hole doping or the tolerance factor, is still lacking.

For the present YBCO/NCSMO superlattice, the hole doping and the tolerance factor of the NCSMO layers are chosen such that they are very close to a phase transition between a FM metallic state (at lower hole doping and high magnetic field) and a predominantly AF and charge/orbital ordered insulating state (toward higher hole doping and low magnetic field). As was shown in Fig. 2 of Ref.(5), and is confirmed by the Mn-XMCD spectra in TEY mode in Fig. S2(f), the NCSMO layers of the present superlattice exhibit a sizeable FM Mn moment with very low field of saturation (around 0.5 T). This behavior is typical for soft ferromagnets and it excludes a possible origin of the manganite FM moment from strong canting of the AF Mn moments. Whether this FM Mn moment arises from a phase separation into AF and FM regions or from a strong alteration of the magnetic properties of the $MnO_2$ layers in the vicinity of the cuprate/manganite interface is not established yet.

## S2. Normalization, self-absorption correction, and fitting of the RIXS spectra

### *S2.1. Normalization and self-absorption correction of the raw data*

The raw counts obtained from the detector were first normalized to the input drain current (which is a measure of incident x-ray flux). The data after this normalization are presented in Fig S4.

Subsequently, a self-absorption correction has been performed that accounts for the change of the interaction volume between the x-ray and the sample (film) that occurs as the Q value respectively the incidence angle of the x-rays is varied. For this self-absorption correction, which plays an important role in the intensities of the inelastic features, we have followed the procedure described in ref.(6, 7). As an example, Fig S5.a shows a comparison of an initial and a self-absorption corrected RIXS spectrum along $[h, 0]$ at 0.33a*.

Finally, to correct for the different footprints of the incident x-ray beam at different incidence angles, the self-absorption corrected RIXS spectra have been normalized to the area of the *dd*-excitations as shown in Fig S5.b. Note that the area of the *dd*-excitation is proportional to the total scattering cross-section of the sample with the incident x-ray which does not depend on the incidence angle.

### *S2.2. Fitting of the normalized data.*

The normalized and self-absorption corrected spectra (as described in section S2.1) have been subsequently fitted with a multi-peak model. As shown in Fig S6, the spectra have been reproduced with (i) 3 peaks to account for the strong *dd*-excitations, consistent with e.g. ref (8), (ii) an elastic peak at zero energy-loss that has a width similar to that of the experimental resolution of 42 meV, (iii) two peaks to account for the high energy phonon-modes (the so-called buckling and stretching mode(9)), (iv) two peaks to account for the magnon modes and (v) one weak peak due to a bimagnon. In addition, we have included a very broad background that accounts for an extension of the high-energy charge-transfer peak. Finally, we have included an additional weak peak around an energy loss of –0.7 eV which arises from the orbital reconstruction effect of the interfacial $CuO_2$ layer that shifts the $d_{x^2-y^2}$.

The magnon bands were fitted with the following equations:

For E$< -0.001$,

$$\tilde{S} = 2E_0\tilde{\chi}\tilde{\Gamma}\frac{E}{\left[1-e^{\frac{E}{k_BT}}\right]\left[(E^2-E_0^2)^2+\left(E\tilde{\Gamma}\right)^2\right]} \qquad \text{(Eq. S2.1)}$$

Otherwise,

$$\tilde{S} = 2E_0\tilde{\chi}\tilde{\Gamma}\frac{E}{\left[1-e^{\frac{E-0.001}{k_BT}}\right]\left[(E^2-E_0^2)^2+\left(E\tilde{\Gamma}\right)^2\right]} \qquad \text{(Eq. S2.2)}$$

Here $\tilde{S}$ is the spectral response of the magnons; $E_0$ the center energy, $\tilde{\chi}$ *the* amplitude and $\tilde{\Gamma}$ the width of the magnon band, and $E$ the energy loss. A more detailed motivation and description of these fit functions can be found in reference (10). The elastic peak and phonons were fitted with the following functions:

$$y = y0 + A_{el}e^{-\frac{(E-E_{el})^2}{W_{el}^2}} \qquad \text{(Eq.}S2.3)$$

$$y = y0 + A_{ph}e^{-\frac{\left(E-E_{ph}\right)^2}{W_{ph}^2}} \qquad \text{(Eq.}S2.4)$$

To account for their skewed shape, the crystal field excitations were fitted with bi-Gaussian functions:

for $x < x_c$

$$y = y0 + He^{-\frac{(x-x_c)^2}{2w_1^2}} \qquad \text{(Eq.}S2.5)$$

else

$$y = y0 + He^{-\frac{(x-x_c)^2}{2w_2^2}} \qquad \text{(Eq.}S2.6)$$

To describe the bimagnons and a weak peak around -0.7eV, that arises from a low-energy crystal field excitation that involves the $d_{3z^2-r^2}$ levels of the interfacial Cu ions that are much closer to the Fermi-level than in the bulk-like Cu ions, we used the following Gaussian function:

$$y = y0 + \left(\frac{A}{w\sqrt{\left(\frac{\pi}{2}\right)}}\right)e^{-2\left(\frac{x-xc}{w}\right)^2} \qquad \text{(Eq.}S2.7)$$

### *S2.3. Calculation of error bars in magnon-parameters*

The error bars in Fig. 4 of the main manuscript have been estimated as follows:

1. Error in position and width: The maximum error in position and width has been estimated to be the resolution of the spectrometer during the experiment: 42 meV.
2. Error in the Area of the magnon peaks: The Area of the magnon peaks can be estimated as:

$$Area_{M1,M2} = (constant) \times height\ of\ the\ peak\ (ht) \times withdth\ of\ the\ peak(w)$$

Consequently, the error in area is estimated as:

$$\delta\left(Area_{M1,M2}\right) = \left|Area_{M1,M2}\right| \times \sqrt{\left(\frac{\delta ht}{ht}\right)^2 + \left(\frac{\delta w}{w}\right)^2}$$

Here, $\delta ht = RMSD = \sqrt{\frac{1}{N}\sum_N (Fit - Exp.data)^2}$ and $\delta w$=42 meV

It is worth mentioning that the error bars estimated here provide the maximum value of error instead of an expectation of error. Following the supplementary info in ref.(11), one can assume that the error in position and width after a successful fit is approximately 10 times smaller than the instrumental resolution. However, since the fitting function for magnons in our analysis is much more complex, we prefer to use the maximum limit of the errors in estimating the error bars.

## S3. Polarization analysis of the RIXS intensity and error bar calculation

### *S3.1. Polarization dependence of scattered x-rays:*

As sketched in Fig. 2A, in a polarimetry experiment one measures (with σ- and π-polarized incident photons) the RIXS intensities after reflection from of a multilayer mirror, $I_{pol}$, and references it to the scattered intensity of the direct beam without the multilayer mirror, $I_{dir}$. With the predetermined values of the polarization-dependent reflectivity of the multilayer-mirror (of $R_\pi = 0.086 \pm 0.002, and\ R_\sigma = 0.141 \pm 0.002$ at Cu L3 edge), one can thus deduce the $I_\pi -$ and $I_\sigma -$components of the RIXS signal according to the following relationship:

$$I_\pi = \frac{R_\sigma I_{dir} - I_{pol}}{R_\sigma - R_\pi}; \qquad I_\sigma = \frac{I_{pol} - R_\pi I_{dir}}{R_\sigma - R_\pi} \qquad \text{(Eq. S3.1)}$$

### *S3.2. Error bar calculation:*

The RIXS intensity has been obtained by normalizing the scattered intensity with the input drain current (proportional to the incident x-ray flux) measured from a reference mirror $M_j$ that is placed on the input side of the spectrometer:

$$I_{RIXS} = \sum_j \frac{I_j}{M_j} \qquad \text{(Eq. S3.2)}$$

For simplicity it is assumed that $M_j$ is almost constant and thus can be approximated by the averaged mirror current M, such that the error of the RIXS intensity can be calculated as:

$$\Delta I_{RIXS} = \frac{1}{M}\sqrt{\sum I_j} \qquad \text{(Eq. S3.3)}$$

We typically made sure that the counts per channel in the single photon counter (SPC) is greater than 20 photons so the distribution can be approximated with a gaussian function for which $\Delta I_{dir,pol} \approx \sqrt{I_{dir,pol}}$. Accordingly, the error bars for the scattered intensities can be calculated using the following standard error propagation:

$$(\Delta I_\pi)^2 \approx \left(\frac{\partial I_\pi}{\partial I_{dir}}\right)^2 (\Delta I_{dir})^2 + \left(\frac{\partial I_\pi}{\partial I_{pol}}\right)^2 (\Delta I_{pol})^2 + \left(\frac{\partial I_\pi}{\partial R_\pi}\right)^2 (\Delta R_\pi)^2 + \left(\frac{\partial I_\pi}{\partial R_\sigma}\right)^2 (\Delta R_\sigma)^2 +$$

$$= \left(\frac{R_\sigma}{R_\sigma - R_\pi}\right)^2 (\Delta I_{dir})^2 + \left(\frac{-1}{R_\sigma - R_\pi}\right)^2 (\Delta I_{pol})^2 + \left(\frac{R_\sigma I_{dir} - I_{pol}}{(R_\sigma - R_\pi)^2}\right)^2 (\Delta R_\pi)^2$$

$$+ \left(\frac{I_{dir}}{R_\sigma - R_\pi} - \frac{R_\sigma I_{dir} - I_{pol}}{(R_\sigma - R_\pi)^2}\right)^2 (\Delta R_\sigma)^2$$

$$(\Delta I_\sigma)^2 \approx \left(\frac{-R_\pi}{R_\sigma - R_\pi}\right)^2 (\Delta I_{dir})^2 + \left(\frac{1}{R_\sigma - R_\pi}\right)^2 (\Delta I_{pol})^2 +$$

$$\left(\frac{I_{pol} - R_\pi I_{dir}}{(R_\sigma - R_\pi)^2} - \frac{I_{dir}}{R_\sigma - R_\pi}\right)^2 (\Delta R_\pi)^2 + \left(\frac{R_\sigma I_{dir} - I_{pol}}{(R_\sigma - R_\pi)^2}\right)^2 (\Delta R_\sigma)^2 \qquad \text{(Eq. S3.4)}$$

## S4. Theoretical estimate of magnon dispersion and RIXS intensity

### *S4.1. Description of the linear spin wave model and the AF ground state.*

We apply a minimal linear spin wave model with a special contribution from the interfacial $CuO_2$ layer to obtain a description of the two magnon modes (M1 and M2) that are observed in the RIXS spectra. Specifically, we adopt a 2D Heisenberg nearest neighbor model for the $CuO_2$ double layers, where only the interfacial $CuO_2$ layer is affected by a weak, antiferromagnetic (AF) exchange interaction, $J_{MC}$, with the Mn moments on the other side of the interface (the latter are assumed to exhibit a soft ferromagnetic order). In addition, we consider that the charge transfer and the related orbital reconstruction at the interface leads to a strong decrease in the population of the $d_{x^2-y^2}$ orbitals and a corresponding increase of the $d_{3z^2-r^2}$ orbitals for the interfacial $CuO_2$ layer *(12)*. Accordingly, we allow for a reduced in-plane AF exchange $J_{\parallel}^{IF}$ of the interfacial $CuO_2$ plane**.** For the other $CuO_2$ plane of this interfacial bilayer-unit, our minimal model assumes that $J_{\parallel}$ recovers already a bulk-like value. This is to reduce the number of fit parameters and motivated by our finding that the RIXS spectra show no sign of a third magnon peak. For the interplanar antiferromagnetic exchange $J_\perp$ between the adjacent $CuO_2$ layers of these bilayer units, we also adopt a bulk-like value.

A sketch of our model is depicted in Fig. 2B of the main text. Here we assume a twinned order of the orthorhombic structure of fully oxygenated and superconducting YBCO described by the space group *Pmmm (13)* with the lattice constants *a=3.82* Å, *b*=3.88Å, *c*=11.68Å. The long range magnetic order is described by the propagation vector ***k*** = (1/2,1/2,0) *(14)* such that the primitive magnetic unit cell with the lattice constants **A** = **a+b**; **B= a - b**; **C=c** contains four magnetic copper ions with coordinates Cu1 =(0,0,z), Cu2 = (1/2, ½, -z), Cu3 (0,0,-z); and Cu4 = (1/2,1/2, z) with z = 0.355(8). The Cu1 and Cu4 ions belong to the interfacial $CuO_2$ plane. It is supposed that the Neel antiferromagnetic vector of the copper bilayer is directed along the $\boldsymbol{a}$ – axis and the ferromagnetic moment of the manganite interfacial plane is directed along the $\boldsymbol{b}$ -axis. The external magnetic field is applied along the direction of the ferromagnetic manganite moment.

The respective Hamiltonian of our model in exchange approximation has the form:

$$\widehat{H}^{(ex)} = \sum_{n,m}\left[J_{\parallel}(\boldsymbol{s}_{2n}\boldsymbol{s}_{3m}) + J_{\parallel}^{IF}(\boldsymbol{s}_{1n}\boldsymbol{s}_{4m})\right] + \sum_{n} J_{\perp}(\boldsymbol{s}_{1n}\boldsymbol{s}_{3n} + \boldsymbol{s}_{4n}\boldsymbol{s}_{2n}) + \\ + \sum_{n} J_{MC}\boldsymbol{n}_{Mn}(\boldsymbol{s}_{1n} + \boldsymbol{s}_{4n}) - \sum_{n} g\mu_B \boldsymbol{H}(\boldsymbol{s}_{1n} + \boldsymbol{s}_{2n} + \boldsymbol{s}_{3n} + \boldsymbol{s}_{4n}) \qquad \text{(Eq. S4.1)}$$

Here $J_{\parallel}$, $J_{\perp}$, $J_{\parallel}^{IF}$, $J_{MC} > 0$. In the following calculation, we assume that the vector $\boldsymbol{n}_{\mathrm{Mn}}$ = (0,1,0) is directed along the manganite's ferromagnetic moment, such that the value of $J_{MC}$ includes the spin of the Mn ions at the interface.

Further, to make the symmetry properties of the calculated results more transparent, we introduce the following linear combinations of the Fourier components of the sublattice spins:

$$\begin{aligned} \boldsymbol{F}(\boldsymbol{k}) &= \boldsymbol{s}_1(\boldsymbol{k}) + \boldsymbol{s}_2(\boldsymbol{k}) + \boldsymbol{s}_3(\boldsymbol{k}) + \boldsymbol{s}_4(\boldsymbol{k}); \\ \boldsymbol{L}_1(\boldsymbol{k}) &= \boldsymbol{s}_1(\boldsymbol{k}) + \boldsymbol{s}_2(\boldsymbol{k}) - \boldsymbol{s}_3(\boldsymbol{k}) - \boldsymbol{s}_4(\boldsymbol{k}); \\ \boldsymbol{L}_2(\boldsymbol{k}) &= \boldsymbol{s}_1(\boldsymbol{k}) - \boldsymbol{s}_2(\boldsymbol{k}) + \boldsymbol{s}_3(\boldsymbol{k}) - \boldsymbol{s}_4(\boldsymbol{k}); \\ \boldsymbol{L}_3(\boldsymbol{k}) &= \boldsymbol{s}_1(\boldsymbol{k}) - \boldsymbol{s}_2(\boldsymbol{k}) - \boldsymbol{s}_3(\boldsymbol{k}) + \boldsymbol{s}_4(\boldsymbol{k}). \end{aligned} \qquad \text{(Eq. S4.2)}$$

Rewriting Hamiltonian (Eq. S5.1) in terms of linear combinations (Eq. S5.2), we get:

$$\begin{aligned} \widehat{H}^{(ex)} = \Sigma\{&A(\boldsymbol{k})[\boldsymbol{F}(\boldsymbol{k})\boldsymbol{F}(-\boldsymbol{k}) - \boldsymbol{L}_1(\boldsymbol{k})\boldsymbol{L}_1(-\boldsymbol{k})] + B(\boldsymbol{k})[\boldsymbol{L}_2(\boldsymbol{k})\boldsymbol{L}_2(-\boldsymbol{k}) \\ &- \boldsymbol{L}_3(\boldsymbol{k})\boldsymbol{L}_3(-\boldsymbol{k})] + C(\boldsymbol{k})[\boldsymbol{F}(\boldsymbol{k})\boldsymbol{L}_3(-\boldsymbol{k}) - \boldsymbol{L}_1(\boldsymbol{k})\boldsymbol{L}_2(-\boldsymbol{k})]\} \\ &+ \frac{1}{2}\sqrt{N}J_{MC}[F_y(0) + L_{3y}(0)] - \sqrt{N}g\mu_B H F_y(0); \end{aligned} \qquad \text{(Eq. S4.3)}$$

Here N is the number of the unit cells in the crystals and the *A, B, and C* coefficients are defined as:

$$\begin{aligned} A(\boldsymbol{k}) &= \frac{1}{8}\left[J_{\perp}(\boldsymbol{k}) + \frac{1}{2}J_{\parallel}(\boldsymbol{k}) + \frac{1}{2}J_{\parallel}^{IF}(\boldsymbol{k})\right]; \\ B(\boldsymbol{k}) &= \frac{1}{8}\left[J_{\perp}(\boldsymbol{k}) - \frac{1}{2}J_{\parallel}(\boldsymbol{k}) - \frac{1}{2}J_{\parallel}^{IF}(\boldsymbol{k})\right]; \\ C(\boldsymbol{k}) &= \frac{1}{8}[J_{\parallel}(\boldsymbol{k}) - J_{\parallel}^{IF}(\boldsymbol{k})]; \end{aligned} \qquad \text{(Eq. S4.4)}$$

with

$J_{\parallel}(\boldsymbol{k}) = 2J_{\parallel}[cos\,\boldsymbol{k\,a} + cos\,\boldsymbol{k\,b}]$;   $J_{\parallel}^{IF}(\boldsymbol{k}) = 2J_{\parallel}^{IF}[cos\,\boldsymbol{k\,a} + cos\,\boldsymbol{k\,b}]$;
$J_{\perp}(\boldsymbol{k}) = J_{\perp}\,cos\,\boldsymbol{k}_{\perp}\,\boldsymbol{d}$;

In our model the Neel vector (magnetic order parameter) of the bulk-like YBCO is directed along the $\boldsymbol{a}$ – axis. Thus, the magnetic ground state is described by the nonzero Neel vector $\bar{L}_{1x} = 4S$ where $S$ is the spin of the copper ions. The magnetic excitations (with four branches of spin waves) of our double layer model for bulk YBCO are described by the set of operators S4.2. In particular, the spin operators $L_{1y}$, $F_z$ and $L_{1z}$, $F_y$ determine the deviation of the primary order parameter from its equilibrium state. In the absence of anisotropy, the corresponding spin waves are gapless Goldstone modes (AM - acoustic modes). For small wave vectors $\boldsymbol{k}$, the AM1 ($L_{1y}$ , $F_z$ ) and AM2 ($L_{1z}$ , $F_y$ ) modes are mainly in-plane ($L_{1y}$) and out-of-plane ($L_{1z}$ ) high-amplitude fluctuations of the Neel vector, respectively. In the exchange approximation at $\boldsymbol{k}$ = 0 their amplitudes diverge. Their fluctuations become less pronounced with increasing $\boldsymbol{k}$.

Another type of spin operators is due to the gap magnetic excitations that break a given type of exchange magnetic order, which therefore requires some exchange energy (OM - optical modes). In our double layer model, the respective gap at $\boldsymbol{k}$=0 is proportional to $\varepsilon_{OM}(0) \propto \sqrt{J_{\perp}J_{\parallel}}$ *(14)*. The sets of operators $L_{3y}$, $L_{2z}$ and $L_{2y}$, $L_{3z}$ describe the OM1 and OM2 optic modes respectively. In the exchange approximation for bulk YBCO ($J_{MC} = 0; J_{\parallel}^{IF} = J_{\parallel}; H = 0$,), the acoustic and optic spin waves do not mix with each other at any $\boldsymbol{k}$ $\neq 0$ in the magnetic Brillouin zone.

The reduced intraplanar exchange $J_{\parallel}^{IF} < J_{\parallel}$, as well the effects of the exchange with the Mn moments, $J_{MC}$, and with the external magnetic field (applied along the b-axis) all tend to decrease the initial magnetic symmetry and give rise to a mixing of the AM1 and OM2 and the AM2 and OM1 modes. However, the main mixing effect arises from the strong reduction of the intraplanar AFM exchange of the interfacial $CuO_2$ layer $J_{\parallel} - J_{\parallel}^{IF}$ that turns out to have by far the largest magnitude of about 60 meV. The respective terms in the Hamiltonian S4.3 are proportional to the coefficient $C(\boldsymbol{k})$.

The magnetic ground state changes according to the new symmetry. Static nonzero components $\bar{F}_y$, $\bar{L}_{1z}$, $\bar{L}_{3y}$, $\bar{L}_{2z}$, $\bar{L}_{2x}$ arise now due to the symmetry breaking by the interfacial manganite layer. Here the y- and z-components describe complex in-plane and out-of-plane tilts of the initial magnetic structure. In the following, we neglect the out-of-plane tilts as they arise as secondary effects caused by the Dzyaloshinskii-Moriya interaction, $D_1F_yL_{1z} + D_2L_{3y}L_{2z}$, which we do not consider in our Heisenberg model. The remaining components $\bar{F}_y$ *and* $\bar{L}_{3y}$ determine the angles of the tilts from the x- towards the y-axis: $\theta_1$ – for the magnetic sublattices 1 and 4 of the interfacial $CuO_2$ plane and $\theta_2$ – for the magnetic sublattices 2 and 3 of the second $CuO_2$ of that bilayer unit. The minimization of the energy thus yields the following relationships:

$$\frac{1}{2}(\bar{F}_y + \bar{L}_{3y}) = \bar{S}_{1y} + \bar{S}_{4y} = 2S\sin\theta_1 = \frac{-J_{MC} + g\mu_B H}{4(J_{\parallel}^{IF} + \frac{J_{\perp}}{8})};$$
$$\frac{1}{2}(\bar{F}_y - \bar{L}_{3y}) = \bar{S}_{2y} + \bar{S}_{3y} = 2S\sin\theta_2 = \frac{g\mu_B H - J_{\perp}S\sin\theta_1}{4J_{\parallel}}; \quad \text{(Eq. S.4.5)}$$

In zero magnetic field, H, the induced magnetic moment of the interfacial Cu ions has a magnitude, $m_{in} = \frac{\mu_B \sin\theta_1}{Cu}$ and is antiparallel to the direction of the ferromagnetic manganite moment. As an external magnetic field is applied, the interfacial Cu moment eventually changes sign at the critical external field, $g\mu_B H_c = J_{MC}$. Our experimental XMCD data in Fig S2 in the main text yield a rough estimate of $H_c \approx 4$ T and thus $J_{MC} \approx 0.5$meV. According to Eq. S4.5, the reduction of $J_{\parallel}^{IF}$ enhances the magnetization of the interfacial $CuO_2$ layer.

The expected tilting angle $\theta_2$ of the Cu2 and Cu3 sublattices in the second copper plane is expected to be opposite and much less than $\theta_1$, since the action of the manganite ferromagnetic moment transfers from the interfacial copper plane to the second copper plane through the AFM interplane exchange $J_{\perp} << J_{\parallel}$. The respective quantitative result is shown in Eq. S4.5. In the following calculations we neglect this weak tilting in the magnetic sublattices 2 and 3 of the second copper plane.

The magnetic moments on the copper ions induced by the external magnetic field in the bulk layers ($J_{MC} = 0; J_{\parallel}^{IF} = J_{\parallel}$) can be defined by the relation:

$$m_b(Cu) = \mu_B \sin\theta = \frac{g\mu_B H}{4(J_{\parallel} + \frac{J_{\perp}}{8})}\mu_B$$

Notice that this result is valid for any direction of the field since it was obtained in the exchange approximation.

### *S4.2. Spin wave calculations*

We use a Holstein - Primakoff approach with a transformation of the sublattice spin operators to the bosonic creation and destruction operators $a_{\alpha}^{+}(\boldsymbol{k}),\ a_{\alpha}(\boldsymbol{k})$ where α =1,2,3,4 are the sublattices numbers. As a first step, we express the sublattice spin operators $s_{i\alpha}(\boldsymbol{k})$ in the crystal coordinate frame with the spin operators $s'_{i\alpha}(\boldsymbol{k})$ in a local coordinate frame for which the z'- axis is directed along the equilibrium value of the spin $\bar{s}_{\alpha}$. We use the relations $s_{i\alpha}(\boldsymbol{k}) = p_{ij}^{(\alpha)} s'_{j\alpha}(\boldsymbol{k})$, where the matrices $\hat{p}^{(\alpha)}$ have the following form:

$$\hat{p}^{(1)} = \begin{pmatrix} \sin\theta & 0 & \cos\theta \\ \cos\theta & 0 & -\sin\theta \\ 0 & 1 & 0 \end{pmatrix}; \quad \hat{p}^{(4)} = \begin{pmatrix} -\sin\theta & 0 & -\cos\theta \\ \cos\theta & 0 & -\sin\theta \\ 0 & -1 & 0 \end{pmatrix}; \quad \text{(Eq. S4.6)}$$

$$\hat{p}^{(2)} = \begin{pmatrix} 0 & 0 & 1 \\ 1 & 0 & 0 \\ 0 & 1 & 0 \end{pmatrix}; \quad \hat{p}^{(3)} = \begin{pmatrix} 0 & 0 & -1 \\ 1 & 0 & 0 \\ 0 & -1 & 0 \end{pmatrix};$$

Here $\theta = \theta_1$ and $\theta_2 = 0$. The local axial y' - axes are always directed along the z-axis of the crystal coordinate frame. With the help of Eq. S4.6 one can construct the following relations between the set of operators $L$ in the form (Eq. S4.2) with operators $L'$ in the same form but with spin operators $s'_{i\alpha}(\boldsymbol{k})$ from the local coordinate frame:

$$\begin{aligned}
F_x(\boldsymbol{k}) &= \frac{1}{2}\{\sin\theta\,(L'_{1x}(\boldsymbol{k}) + L'_{2x}(\boldsymbol{k})) + L'_{1z}(\boldsymbol{k})(\cos\theta + 1) + L'_{2z}(\boldsymbol{k})(\cos\theta - 1)\}; \\
F_y(\boldsymbol{k}) &= \frac{1}{2}\{(\cos\theta + 1)F'_x(\boldsymbol{k}) + (\cos\theta - 1)L'_{3x}(\boldsymbol{k}) - \sin\theta\,(F'_z(\boldsymbol{k}) + L'_{3z}(\boldsymbol{k}))\}; \\
L_{1x}(\boldsymbol{k}) &= \frac{1}{2}\{\sin\theta\,(F'_x(\boldsymbol{k}) + L'_{3x}(\boldsymbol{k})) + F'_z(\boldsymbol{k})(\cos\theta + 1) + L'_{3z}(\boldsymbol{k})(\cos\theta - 1)\}; \\
L_{1y}(\boldsymbol{k}) &= \frac{1}{2}\{(\cos\theta + 1)L'_{1x}(\boldsymbol{k}) + (\cos\theta - 1)L'_{2x}(\boldsymbol{k}) - \sin\theta\,(L'_{1z}(\boldsymbol{k}) + L'_{2z}(\boldsymbol{k}))\}; \\
L_{2x}(\boldsymbol{k}) &= \frac{1}{2}\{\sin\theta\,(F'_x(\boldsymbol{k}) + L'_{3x}(\boldsymbol{k})) + F'_z(\boldsymbol{k})(\cos\theta - 1) + L'_{3z}(\boldsymbol{k})(\cos\theta + 1)\}; \\
L_{2y}(\boldsymbol{k}) &= \frac{1}{2}\{(\cos\theta - 1)L'_{1x}(\boldsymbol{k}) + (\cos\theta + 1)L'_{2x}(\boldsymbol{k}) - \sin\theta\,(L'_{1z}(\boldsymbol{k}) + L'_{2z}(\boldsymbol{k}))\}; \\
L_{3x}(\boldsymbol{k}) &= \frac{1}{2}\{\sin\theta\,(L'_{1x}(\boldsymbol{k}) + L'_{2x}(\boldsymbol{k})) + L'_{1z}(\boldsymbol{k})(\cos\theta - 1) + L'_{2z}(\boldsymbol{k})(\cos\theta + 1)\}; \\
L_{3y}(\boldsymbol{k}) &= \frac{1}{2}\{(\cos\theta - 1)F'_x(\boldsymbol{k}) + (\cos\theta + 1)L'_{3x}(\boldsymbol{k}) - \sin\theta\,(F'_z(\boldsymbol{k}) + L'_{3z}(\boldsymbol{k}))\}; \\
F_z(\boldsymbol{k}) &= L'_{1y}(\boldsymbol{k}); \quad L_{1z}(\boldsymbol{k}) = F'_y(\boldsymbol{k}); \quad L_{2z}(\boldsymbol{k}) = L'_{3y}(\boldsymbol{k}); \quad L_{3z}(\boldsymbol{k}) = L'_{2y}(\boldsymbol{k});
\end{aligned} \qquad \text{(Eq S4.7)}$$

To make the symmetry properties more transparent, we introduce the same linear combinations (Eq. S4.2) for the sublattice bosonic operators $a_\alpha(\boldsymbol{k})$, for $\alpha = 1,2,3,4$ (12):

$$a_F(\boldsymbol{k}) = \frac{1}{2}\{a_1(\boldsymbol{k}) + a_2(\boldsymbol{k}) + a_3(\boldsymbol{k}) + a_4(\boldsymbol{k})\};$$

$$a_{L_1}(\boldsymbol{k}) = \frac{1}{2}\{a_1(\boldsymbol{k}) + a_2(\boldsymbol{k}) - a_3(\boldsymbol{k}) - a_4(\boldsymbol{k})\};$$

$$a_{L_2}(\boldsymbol{k}) = \frac{1}{2}\{a_1(\boldsymbol{k}) - a_2(\boldsymbol{k}) + a_3(\boldsymbol{k}) - a_4(\boldsymbol{k})\}; \qquad \text{(Eq S4.8)}$$

$$a_{L_3}(\boldsymbol{k}) = \frac{1}{2}\{a_1(\boldsymbol{k}) - a_2(\boldsymbol{k}) - a_3(\boldsymbol{k}) + a_4(\boldsymbol{k})\}.$$

Next, we apply the Holstein-Primakoff transformation to convert the operators ***L'*** to the bosonic operators $a_L(\boldsymbol{k}),\ a_L^+(\boldsymbol{k})$ ( *L= F, L₁, L₂, L₃*):

$$L'_x(\boldsymbol{k}) = \sqrt{4s}Q_L(\boldsymbol{k});\ L'_y(\boldsymbol{k}) = i\sqrt{4s}P_L(-\boldsymbol{k});$$

$$Q_L(\boldsymbol{k}) = \frac{1}{\sqrt{2}}\{a_L^+(-\boldsymbol{k}) + a_L(\boldsymbol{k})\};\ P_L(\boldsymbol{k}) = \frac{1}{\sqrt{2}}\{a_L^+(\boldsymbol{k}) - a_L(-\boldsymbol{k})\} \qquad \text{(Eq S4.9)}$$

The z' components of the operators ***L'*** can be expressed in terms of operators *Q* and *P* (see Eq.10 in ref *15*). With the help of the relations (Eq. S4.7) and (Eq. S4.9) we can thus obtain a spin-wave Hamiltonian that is the quadratic part of the Hamiltonian (Eq. S4.3) with respect to bosonic operators (Eq. S4.8):

$$\widehat{H}^{(2)} = \frac{1}{2}\sum_{k,\sigma}\{q_\sigma(\boldsymbol{k})Q_\sigma(\boldsymbol{k})Q_\sigma(-\boldsymbol{k}) - p_\sigma(\boldsymbol{k})P_\sigma(\boldsymbol{k})P_\sigma(-\boldsymbol{k})\} +$$
$$+\sum_k\left\{\begin{matrix} q_{03}(\boldsymbol{k})Q_0(\boldsymbol{k})Q_3(-\boldsymbol{k}) - p_{03}(\boldsymbol{k})P_{03}(\boldsymbol{k})P_L(-\boldsymbol{k}) + \\ +q_{12}(\boldsymbol{k})Q_1(\boldsymbol{k})Q_2(-\boldsymbol{k}) - p_{12}(\boldsymbol{k})P_1(\boldsymbol{k})P_2(-\boldsymbol{k}) \end{matrix}\right\} \qquad \text{(Eq. S4.10)}$$

Here the subscripts σ=0,1,2,3 correspond to σ=*F, L₁, L₂, L₃*. The coefficients $q_\sigma(\boldsymbol{k})$ and $p_\sigma(\boldsymbol{k})$ in the Hamiltonian (Eq. S4.10) are defined by the expressions:

$$q_0(\boldsymbol{k}) = 8S[a(0) + d(\boldsymbol{k})]; p_0(\boldsymbol{k}) = 8S[a(0) - A(\boldsymbol{k})];$$
$$q_1(\boldsymbol{k}) = 8S[a(0) - d(\boldsymbol{k})];\ p_1(\boldsymbol{k}) = 8S[a(0) + A(\boldsymbol{k})];$$
$$q_2(\boldsymbol{k}) = 8S[a(0) - b(\boldsymbol{k})]; p_2(\boldsymbol{k}) = 8S[a(0) - B(\boldsymbol{k})];$$
$$q_3(\boldsymbol{k}) = 8S[a(0) + b(\boldsymbol{k})];\ p_3(\boldsymbol{k}) = 8S[a(0) + B(\boldsymbol{k})]; \qquad \text{(Eq. S4.11)}$$
$$q_{12}(\boldsymbol{k}) = 4S[C(0) - c(\boldsymbol{k})];\ \ p_{12}(\boldsymbol{k}) = 4S[C(0) + C(\boldsymbol{k})];$$
$$q_{03}(\boldsymbol{k}) = 4S[C(0) + c(\boldsymbol{k})];\ \ p_{03}(\boldsymbol{k}) = 4S[C(0) - C(\boldsymbol{k})];$$

$$a(0) = [(A(0) - B(0))\cos\theta + (A(0) + B(0))(1 + \sin^2\theta)]/2;$$
$$d(\boldsymbol{k}) = [A(\boldsymbol{k})(1 + \cos\theta) + B(\boldsymbol{k})(1 - \cos\theta) - (A(\boldsymbol{k}) - B(\boldsymbol{k}) + C(\boldsymbol{k}))\sin^2\theta]/2;$$

$$b(\boldsymbol{k}) = [A(\boldsymbol{k})(1-\cos\theta) + B(\boldsymbol{k})(1+\cos\theta) - (A(\boldsymbol{k}) - B(\boldsymbol{k}) + C(\boldsymbol{k}))\sin^2\theta]/2;$$
$$c(\boldsymbol{k}) = C(\boldsymbol{k}) - (A(\boldsymbol{k}) - B(\boldsymbol{k}) + C(\boldsymbol{k}))\sin^2\theta\,;$$

The energy of the spin waves will be defined by the standard Bogolyubov transformation of the bosonic operators $a_L(\boldsymbol{k})$, $a_L^+(\boldsymbol{k})$ into the $\xi_\nu^+(\boldsymbol{k})$, $\xi_\nu(\boldsymbol{k})$ operators of the creation and annihilation a magnon of the v$^{th}$ branch:

$$Q_L(\boldsymbol{k}) = \frac{1}{\sqrt{2}}\sum_\nu t_{L\nu}(\boldsymbol{k})\{\xi_\nu^+(-\boldsymbol{k}) + \xi_\nu(\boldsymbol{k})\}$$
$$P_L(\boldsymbol{k}) = \frac{1}{\sqrt{2}}\sum_\nu d_{L\nu}(\boldsymbol{k})\{\xi_\nu^+(\boldsymbol{k}) - \xi_\nu(-\boldsymbol{k})\} \qquad \text{(Eq. S4.12)}$$

The coefficients *t* and *d* satisfy the usual normalization conditions.

$$\sum_\nu \{t_{L\nu}(\boldsymbol{k})d_{L'\nu}^*(\boldsymbol{k}) + t_{L'\nu}^*(\boldsymbol{k})d_{L\nu}(\boldsymbol{k})\} = 2\delta_{LL'}$$
$$\sum_L \{t_{L\nu}(\boldsymbol{k})d_{L\nu'}^*(\boldsymbol{k}) + t_{L\nu'}^*(\boldsymbol{k})d_{L\nu}(\boldsymbol{k})\} = 2\delta_{\nu\nu'} \qquad \text{(Eq. S4.13)}$$

The pairs of acoustic modes and the pairs of optical modes are degenerate within the accuracy of the exchange approximation.

The spin wave energies and the *t-d* coefficients of bulk YBCO ($J_{MC} = 0; J_\parallel^{IF} = J_\parallel; \theta = 0;$) have the simple form:

$$t_{L\nu}(\boldsymbol{k}) = \{\frac{p_L(\boldsymbol{k})}{q_L(\boldsymbol{k})}\}^{\frac{1}{4}}; \quad d_{L\nu}(\boldsymbol{k}) = \{\frac{q_L(\boldsymbol{k})}{p_L(\boldsymbol{k})}\}^{\frac{1}{4}};$$
$$\varepsilon_{AM1}(\boldsymbol{k}) = \sqrt{p_0(\boldsymbol{k})q_0(\boldsymbol{k})}; \qquad \varepsilon_{AM2}(\boldsymbol{k}) = \sqrt{p_1(\boldsymbol{k})q_1(\boldsymbol{k})}; \qquad \text{(Eq. S4.14)}$$
$$\varepsilon_{OM1}(\boldsymbol{k}) = \sqrt{p_2(\boldsymbol{k})q_2(\boldsymbol{k})}; \qquad \varepsilon_{OM2}(\boldsymbol{k}) = \sqrt{p_3(\boldsymbol{k})q_3(\boldsymbol{k})};$$

The above-mentioned divergence of the acoustic spin wave fluctuations is evident from the coefficients $d_{0AM1}(\boldsymbol{k}) = \{\frac{q_0(\boldsymbol{k})}{p_0(\boldsymbol{k})}\}^{\frac{1}{4}}$; $t_{1AM2}(\boldsymbol{k}) = \{\frac{p_1(\boldsymbol{k})}{q_1(\boldsymbol{k})}\}^{\frac{1}{4}}$; for which $\lim_{k\to 0} p_0(\boldsymbol{k}) = 0$; $\lim_{k\to 0} q_1(\boldsymbol{k}) = 0$. The relations (Eq. S4.14) for magnon energies can be used to estimate the magnitude of the intraplane exchange interaction of the bulk-like $CuO_2$ layer from the M2 mode that governs the RIXS data at 931 eV.

The fitting, shown in Fig. 5A and 5B in the main manuscript yields a value of $J_{||}$ = 130 meV that is typical for bulk YBCO(*11*, *14*). Here we have used a fixed value of $J_\perp$ = 7 meV that also agrees well with bulk-like properties.

The last term of the summation in the Hamiltonian in equation Eq. S4.10 represents a set of mixing terms between two pairs of acoustic and optic modes that arises from the inequality of the interfacial and bulk-like $CuO_2$ planes that is caused by the reduction of the intraplanar exchange $\delta J_{\parallel} = J_{\parallel} - J_{\parallel}^{IF}$. Note that there exists no corresponding mixing between the pairs of acoustic modes and between the pairs of optical modes which thus remain double degenerated in $\boldsymbol{k}$-space. The respective solutions for the spin wave energies and the t-d coefficients in our double layer model can be found analytically. We apply these solutions to determine the value of $\delta J_{\parallel} = J_{\parallel} - J_{\parallel}^{IF}$ from a fit to the M1 mode in the RIXS data at 930.5 eV as shown in Fig. 5A and 5B.

From the above-described fitting we obtain an unusually large suppression of the interfacial in-plane AFM exchange of $\delta J_{\parallel} = J_{\parallel} - J_{\parallel}^{IF} = 60$ meV to a value of $J_{\parallel}^{IF} = 70$ meV. The energy splitting arises from the mixing between the acoustic and optical spin waves which increases towards larger $\boldsymbol{k}$-vectors and becomes maximal at the Brillouin zone boundary. We recall that this strong reduction of the energy of the interfacial magnon mode (M1-mode) can be naturally explained in terms of the orbital reconstruction of interface copper ions which leads to a mixed $d_{x^2-y^2}$ and $d_{3z^2-r^2}$ orbital character of the holes.

### *S4.3. Intensity of spin waves in the RIXS spectra of YBCO/NCSMO multilayers*

According to Ref. *(16,17)* the RIXS intensity of the spin waves of the $\nu$-th branch in cross-polarization is defined by the square of the matrix element corresponding to the transition between the states with a difference in magnon numbers $n_\nu(\boldsymbol{k})$ equal to unity:

$$M_{k\nu} \propto \langle n_\nu(\boldsymbol{k}) + 1 | (\boldsymbol{\sigma}_{out} \times \boldsymbol{\pi}_{in}) \cdot \{ \sum_{\alpha=1,2,3,4} f_\alpha(\boldsymbol{Q}) e^{-i2\pi \boldsymbol{Q}\boldsymbol{\rho}_\alpha} \boldsymbol{S}_\alpha(\boldsymbol{k}) \} | n_\nu(\boldsymbol{k}) \rangle \quad \text{(Eq. S4.15)}$$

Here $\boldsymbol{\pi}_{in}, \boldsymbol{\sigma}_{out}$ denote the polarizations of the incoming and outgoing x-ray beams, $\boldsymbol{Q}$ is the scattering vector, $f_\alpha(\boldsymbol{Q})$ the form-factor of the copper ion α, and $\boldsymbol{\rho}_\alpha$ the position of copper ion α in the primitive magnetic cell. For the bulk-like YBCO layers equation (Eq. S4.15) can be expressed in terms of operators $\boldsymbol{L}$ using the relation (Eq. S4.2).

$$M_{k\nu} \propto \langle n_\nu(\boldsymbol{k}) + 1 | (\boldsymbol{\sigma_{out}} \times \boldsymbol{\pi_{in}}) \cdot \{ f(\boldsymbol{Q}) \sum_{L=0,1,2,3} G_L(\boldsymbol{Q}) \boldsymbol{L}(\boldsymbol{k}) \} | n_\nu(\boldsymbol{k}) \rangle; \quad \text{(Eq. S4.16)}$$

Here it is supposed that all copper ions are in the same orbital state and therefore have the same form-factor, $f(\boldsymbol{Q})$. The $\boldsymbol{Q}$ – dependence of the form-factor for the copper ions at the Cu $L$-edge can be derived in the frame of a single ion model described in Ref. *(18)* assuming, for example, that all the holes are in a $d_{x^2-y^2}$ orbital ground state. This calculation considers the atomic symmetries of the core-level $2p$ and valence $3d$-orbitals and the geometry of the experimental setup, namely π–polarized incident light and a scattering angle of 50°. The matrix elements for elastic scattering are therefore obtained as the product of two dipole transitions: between the core-level state $2p$ to a valence state $3d$ and de-excitation from the same $3d$ state back to the $2p$ state. Selection rules involving the addition of angular momenta are computed using the Wigner 3-*j* symbols.

As we are interested in the intensity of magnetic excitations close to the elastic line, we calculate cross-sections involving a spin-flip with a spin initially along the $\boldsymbol{a}$ direction for either $d_{x^2-y^2}$ or $d_{3z^2-r^2}$ ground state orbitals (but not involving any change of orbital character in the *3d* valence state).

Fig S8. summarizes the results for the single ion-model calculation. Please note that any linear combination of the data in Fig S8(a) and S8(b) cannot reproduce the experimental trend. It indulges us to account for the $t-d$ coefficients.

The geometrical (or structural) factors $G_L(\boldsymbol{Q})$ are analogous to those for inelastic neutron scattering on spin waves *(15)*:

$$G_0(\boldsymbol{Q}) = cos\frac{\pi}{2}\boldsymbol{Q}(\boldsymbol{A}+\boldsymbol{B})\,cos\,\pi\,\boldsymbol{Q}_\perp\boldsymbol{d};\quad G_1(\boldsymbol{Q}) = -\,sin\frac{\pi}{2}\boldsymbol{Q}(\boldsymbol{A}+\boldsymbol{B})\,sin\,\pi\,\boldsymbol{Q}_\perp\boldsymbol{d}$$

$$G_2(\boldsymbol{Q}) = i\,sin\frac{\pi}{2}\boldsymbol{Q}(\boldsymbol{A}+\boldsymbol{B})\,cos\,\pi\,\boldsymbol{Q}_\perp\boldsymbol{d};\quad G_3(\boldsymbol{Q}) = i\,cos\frac{\pi}{2}\boldsymbol{Q}(\boldsymbol{A}+\boldsymbol{B})\,sin\,\pi\,\boldsymbol{Q}_\perp\boldsymbol{d}; \qquad \text{(Eq. 4.17)}$$

Here we use a bulk-like value for the distance between copper ions along the z-axis of $d$ = 3.388Å. Finally, the $\boldsymbol{Q}$-dependence of the RIXS magnon cross-section can be expressed in terms of the $\boldsymbol{\sigma}_{out}$, $\boldsymbol{\pi}_{in}$-polarization, the form factor $f(\boldsymbol{Q})$, the structural factor $G(\boldsymbol{Q})$ and the *t-d* coefficients of the spin – waves.

For the resonance condition of the interfacial $CuO_2$ layer at 930.5 eV, for which mainly the interfacial copper ions Cu1 and Cu4 ions participate in the scattering process, we assume that the contributions of the Cu2 and Cu3 ions (from the other $CuO_2$ layer of the bilayer unit that is located further away from the interface) can be omitted. The respective matrix element for the RIXS process thus has the following form.

$$M_{k\nu} \propto \langle n_\nu(\boldsymbol{k})+1|(\boldsymbol{\sigma_{out}}\times\ \boldsymbol{\pi_{in}})\cdot f(\boldsymbol{Q})\cdot\{e^{-i2\pi Q\rho_1}\boldsymbol{S}_1(\boldsymbol{k}) + e^{-i2\pi Q\rho_4}\boldsymbol{S}_4(\boldsymbol{k})\}|n_\nu(\boldsymbol{k})\rangle \qquad \text{(Eq. 4.18)}$$

Using the relation (Eq. S4.2) we thus can rewrite the equation (Eq. S4.18) for the interfacial $CuO_2$ layers in terms of the following form of the $\boldsymbol{L}$ operators:

$$M_{k\nu} \propto \frac{1}{2} \langle n_\nu(\boldsymbol{k}) + 1 | (\boldsymbol{\sigma_{out}} \times \quad \boldsymbol{\pi_{in}}) \cdot f(\boldsymbol{Q}) \cdot \{cos \frac{\pi}{2} \boldsymbol{Q}(\boldsymbol{A}+\boldsymbol{B})[\boldsymbol{F}(\boldsymbol{k}) + \boldsymbol{L}_3(\boldsymbol{k})] + i \, sin \frac{\pi}{2} \boldsymbol{Q}(\boldsymbol{A}+\boldsymbol{B})[\boldsymbol{L}_1(\boldsymbol{k}) + \boldsymbol{L}_2(\boldsymbol{k})]\} | n_\nu(\boldsymbol{k}) \rangle \qquad \text{(Eq. S4.19)}$$

In this simplified case, the scattering occurs only from the interfacial $CuO_2$ layer, the contribution of $\boldsymbol{Q}_\perp$ to the structural factor thus is absent. Furthermore, we demonstrate that the doubly degenerate acoustic modes dominate the scattering from the magnons in the interfacial $CuO_2$ layer, whereas the contribution from the optical magnon modes can be neglected.

It is important to note a rather general similarity between Eq. S4.19 and Eq. S4.16 with respect to a divergence of the acoustic spin wave fluctuations at $\boldsymbol{k} \rightarrow 0$ ascribed by operators $L_{1y}$ and $L_{1z}$ that is in both cases fully suppressed by the structural factors. However, theoretically the divergence of the RIXS magnon cross section can occur at the $\boldsymbol{Q}_{||} = \boldsymbol{A}^*$ (at so called (π, π) -point, the center of the next magnetic Brillouin zone) which can be observed if $Q_\perp \neq 0$ for the scattering on the bulk (*17*) and under absence of such a restriction for the scattering on the interfacial plane.

The $\boldsymbol{Q}$ – dependence of the RIXS intensity as calculated for the bulk-like $CuO_2$ planes, with the help of Eq. S4.16, and for the interfacial $CuO_2$ layer, with the help of Eq. S4.19 is shown by the solid lines in Figs. 5C-5F of the manuscript. The calculated intensity contains contributions of a sum of independent contributions for the four spin wave branches (i.e., from two doubly degenerate optic modes and the two doubly degenerate acoustic modes). For the bulk like magnon mode (M2 mode) there is reasonable agreement in Figs. 5C and 5D between the measured RIXS data and the prediction of our minimal spin-wave theory. To the contrary, for the M1 mode due to the magnons in the interfacial $CuO_2$ layers, there a clear discrepancy between the $\boldsymbol{Q}$-dependence of the measured values and the theoretical prediction. Notably, the former exhibit a steep increase of the scattering intensity towards small k-values whereas the latter predict a strong decrease.

The suppression of the contribution of the highly intensive acoustic spin-wave fluctuations to the inelastic scattering process at $\boldsymbol{k}$ vectors close to the Γ – point has a very general origin. Indeed, the acoustic low energy modes induce a rotation of the main order parameter that involves spin fluctuations of equal magnitude but of opposite directions on the nearest neighbor sites. Therefore, irrespective of the number of AFM sublattices, the scattering cross-section of acoustic spin-waves is strongly suppressed and vanishes in the limit of $\boldsymbol{k} \rightarrow 0$, due to the mutual cancelation of in-phase and out-of-phase fluctuations with the same scattering amplitude at the neighboring up- and down- spin sites. This destructive-interference-effect manifests itself as an action of the structural factor that is proportional to $\boldsymbol{k}$, as is evident from Eq. S4.16 and Eq. S4.19.

The obvious way to overcome this destructive interference effect on the magnon intensity towards small wave vectors is a mechanism that gives rise to different scattering amplitudes of the neighboring Cu ions. A natural candidate in the context of the cuprate/manganite interface is a spatial alternation of the orbital character of the holes of the interfacial Cu ions. The simplest model that we consider is the checkerboard type of the $d_{x^2-y^2}$ and $d_{3z^2-r^2}$ orbital order shown in Fig. 6A of the main manuscript which has the same translation symmetry as the underlying AF spin order. Note, that this orbital order also implies a strong decrease of the interfacial intraplane exchange. The resulting scattering amplitude of the (M1) magnon excitations in the RIXS spectra has the form:

$$M_{kv} \propto \langle n_v(\boldsymbol{k}) + 1|(\boldsymbol{\sigma_{out}} \times \boldsymbol{\pi_{in}}) \cdot \{f(\boldsymbol{Q}, x^2)e^{-i2\pi \boldsymbol{Q}\boldsymbol{\rho}_1}\boldsymbol{S}_1(\boldsymbol{k}) + f(\boldsymbol{Q}, z^2)e^{-i2\pi \boldsymbol{Q}\boldsymbol{\rho}_4}\boldsymbol{S}_4(\boldsymbol{k})\}|n_v(\boldsymbol{k})\rangle \quad \text{(Eq. S4.20)}$$

Here we introduce the different form-factors $f(\boldsymbol{Q}, x^2)$, $f(\boldsymbol{Q}, z^2)$ for the copper ions in the $d_{x^2-y^2}$, $d_{3z^2-r^2}$ orbital states, respectively. The different orbital states provide different scattering amplitudes through the different amplitudes of the form-factor. After respective substitutions of the sublattices spin operators $\boldsymbol{S}$ into the operators $\boldsymbol{L}$ we thus obtain:

$$\begin{aligned} M_{kv} \propto \frac{1}{2}\langle n_v(\boldsymbol{k}) + 1|(\boldsymbol{\sigma_{out}} \times \boldsymbol{\pi_{in}}) \cdot \\ \{[f(\boldsymbol{Q}, x^2) + f(\boldsymbol{Q}, z^2)][cos\frac{\pi}{2}\boldsymbol{Q}(\boldsymbol{A} + \boldsymbol{B})[\boldsymbol{F}(\boldsymbol{k}) + \boldsymbol{L}_3(\boldsymbol{k})] + i\, sin\frac{\pi}{2}\boldsymbol{Q}(\boldsymbol{A} + \boldsymbol{B})[\boldsymbol{L}_1(\boldsymbol{k}) + \boldsymbol{L}_2(\boldsymbol{k})]] + \\ +[f(\boldsymbol{Q}, x^2) - f(\boldsymbol{Q}, z^2)][cos\frac{\pi}{2}\boldsymbol{Q}(\boldsymbol{A} + \boldsymbol{B})[\boldsymbol{L}_1(\boldsymbol{k}) + \boldsymbol{L}_2(\boldsymbol{k})] + i\, sin\frac{\pi}{2}\boldsymbol{Q}(\boldsymbol{A} + \boldsymbol{B})[\boldsymbol{F}(\boldsymbol{k}) + \boldsymbol{L}_3(\boldsymbol{k})]]\}|n_v(\boldsymbol{k})\rangle \end{aligned} \quad \text{(Eq. S4.21)}$$

The first term in the braces represents the result described above in Eq. S4.19. Here, at low $\boldsymbol{k}$, the divergent contributions from the operators $L_{1y}$ and $L_{1z}$, are fully suppressed by the structure factor $sin\frac{\pi}{2}\boldsymbol{Q}(\boldsymbol{A} + \boldsymbol{B})$.

In contrast, for the second term in the braces, which is proportional to the form-factor difference $f(\boldsymbol{Q}, x^2) - f(\boldsymbol{Q}, z^2)$, the contribution of the strong spin-wave fluctuations of the magnetic order parameter does not get suppressed toward small wave vectors. It rather gives rise to a contribution of the acoustic magnons of the interfacial $CuO_2$ layers that strongly increases towards small wave vectors and thus can explain the observed strong increase of the M1 mode intensity in the RIXS spectra, as shown by the solid line in Fig. 6B in the main manuscript. Here the different form factors for the spin with $d_{x^2-y^2}$ and $d_{3z^2-r^2}$ orbital character have been calculated with the single ion model described in Ref.(18). Note that the twinning of the orthorhombic structure of YBCO has been considered by averaging their contributions to the RIXS cross-section.

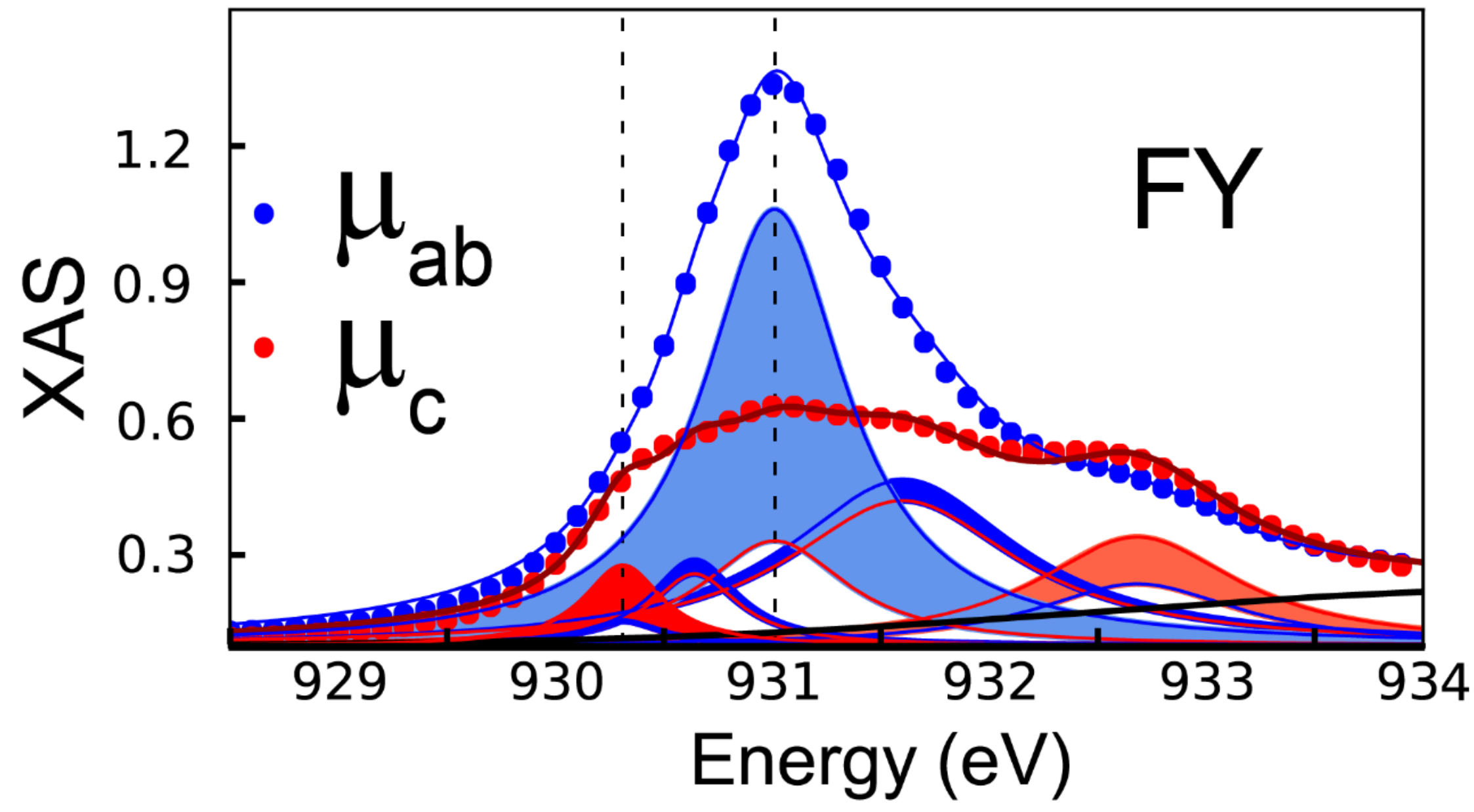

**Fig. S1.** FY spectra and multi-peak fit for the in-plane and out-of-plane components of the linear polarized incident x-rays. Blue (red) shaded areas denote contributions for which $\boldsymbol{\mu_{ab}} - \boldsymbol{\mu_c}$ >0. (<0).

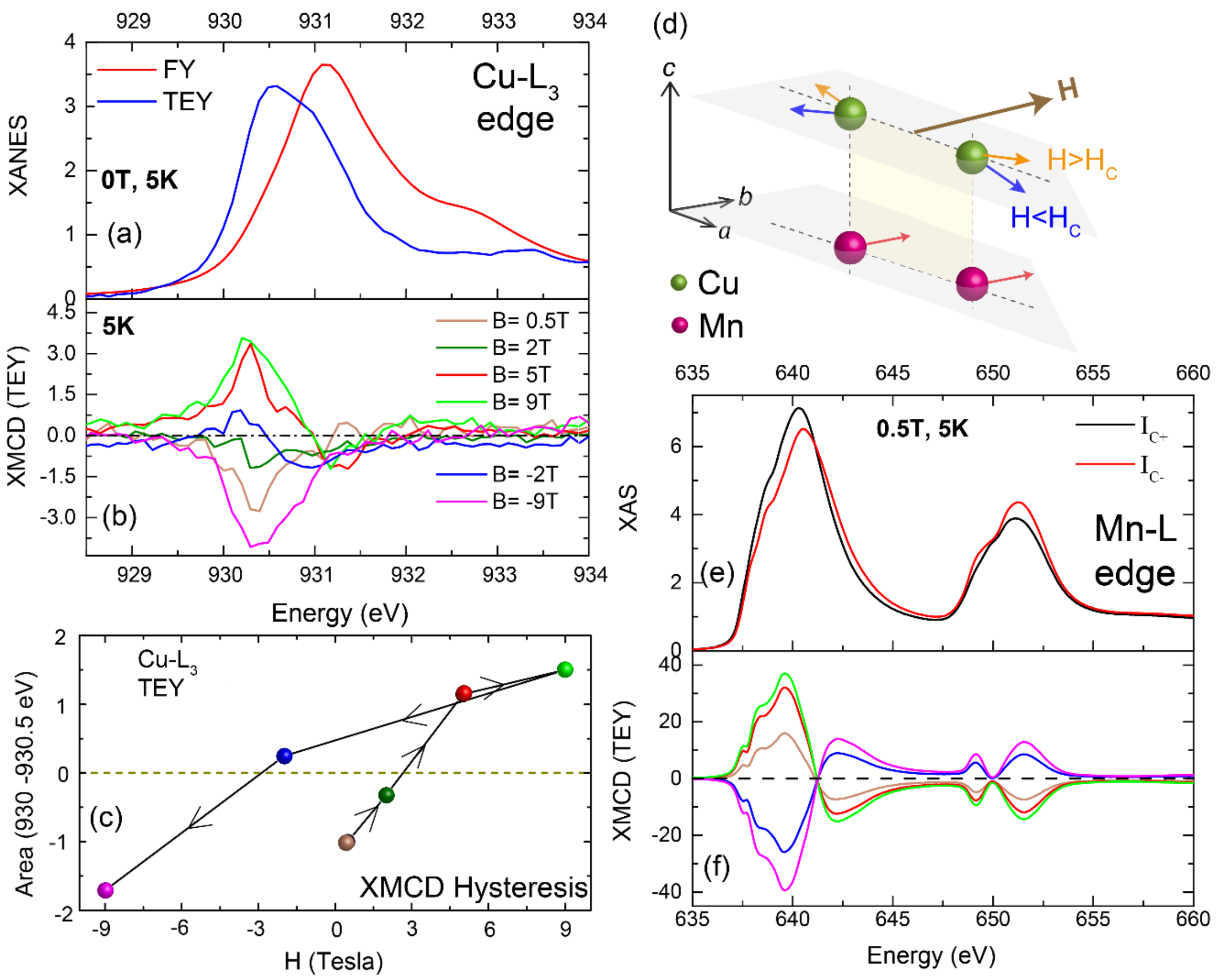

**Fig. S2.** (a) XAS spectra in FY and TEY modes that reveal the distinct resonances of the bulk-like and interfacial Cu ions around 930.4 and 931 eV, respectively. (b) Corresponding Cu-XMCD spectra in TEY mode at different applied fields. They confirm that the magnetic signal is maximal around the resonance of the interfacial Cu ions at 930.4 eV and that it undergoes a sign change around 3 − 4 Tesla that is characteristic of a field-induced reorientation. (c) Magnetic field-loop of the integrated Cu-XCMD signal in panel (b). (d) Sketch of the mutual orientation of the ferromagnetic Mn moment and the induced Cu moment, due to the canting of the AF Cu-spins, below (blue) and above (orange) the switching field, $H_c$. (e) Circularly polarized Mn-XAS spectra in TEY mode. (f) Corresponding Mn-XMCD spectra at different magnetic fields which confirm that the Mn moment is always parallel to the applied field.

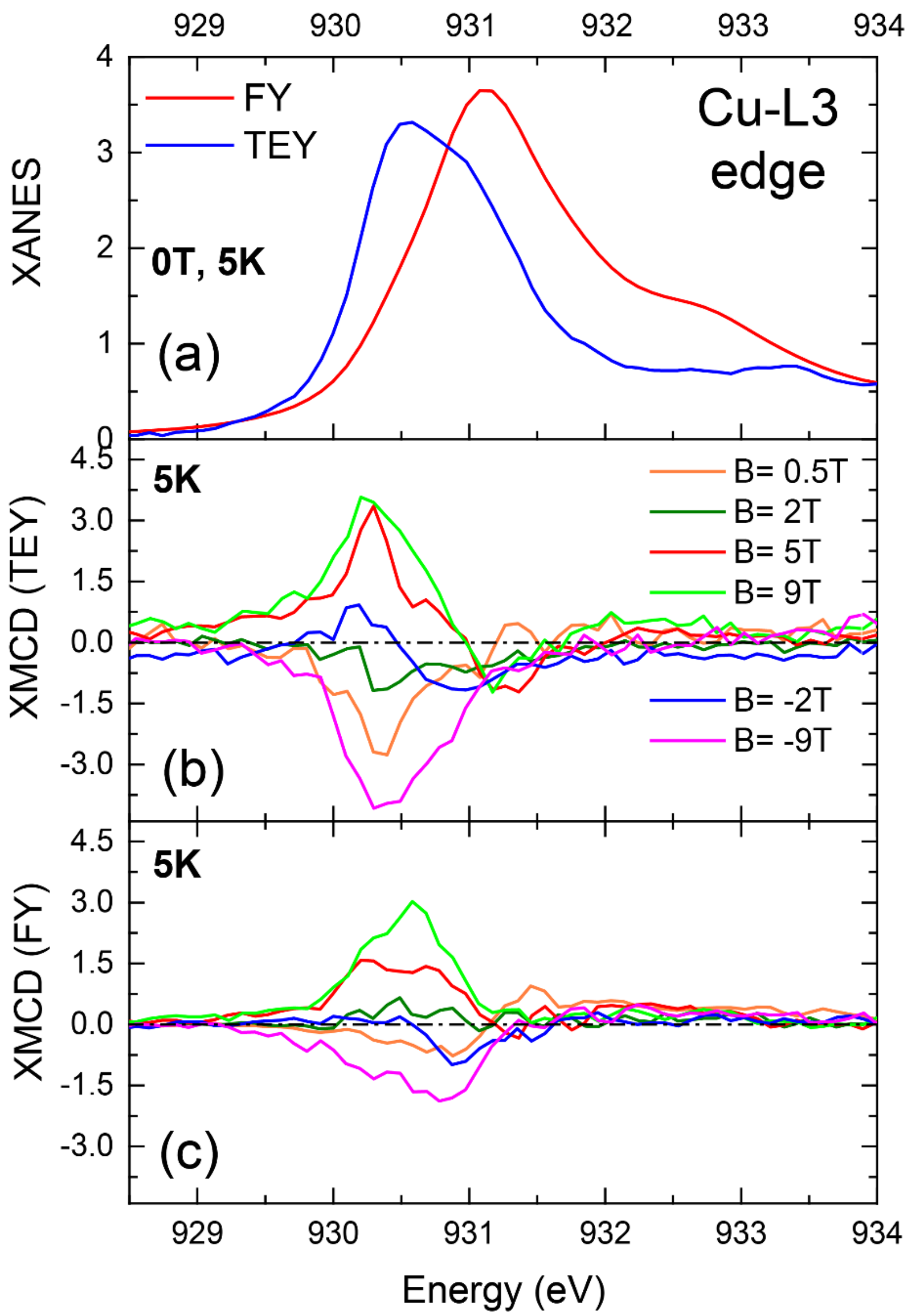

**Fig. S3.** Comparison of the XMCD in FY and TEY spectra at Cu-L3 edge.

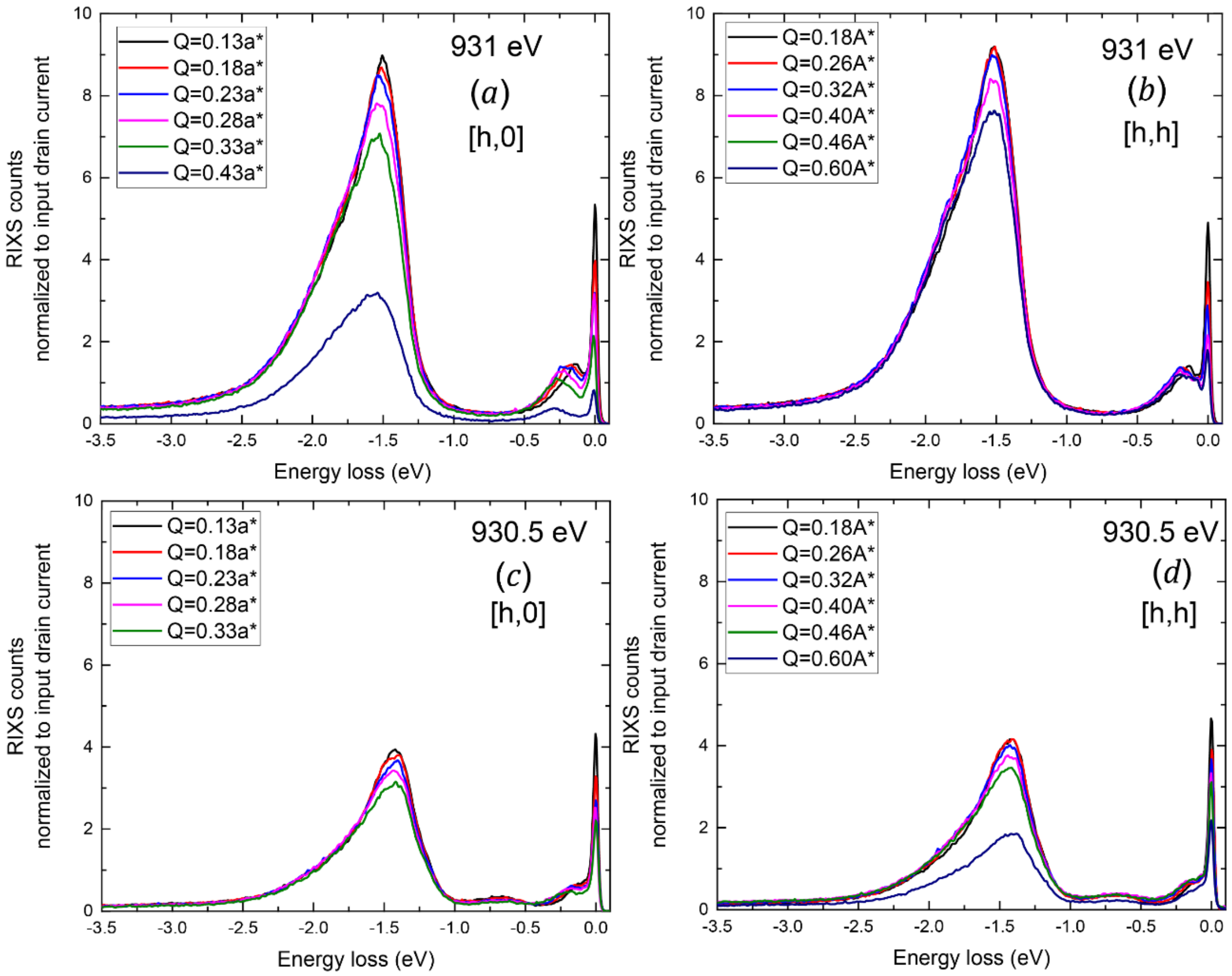

**Fig S4.** Cu-$L_3$ edge RIXS counts obtained by normalizing the signal to the input x-ray flux only: (a) and (b): RIXS spectra taken at the resonance of the bulk-like Cu ions at 931 eV along [h,0] and [h,h], respectively. (c) and (d): Corresponding RIXS spectra taken at the resonance of the interfacial Cu ions at 930.5 eV. The data taken at the largest value of Q|| show an intensity decrease that arise from a larger footprint of the beam at grazing incidence.

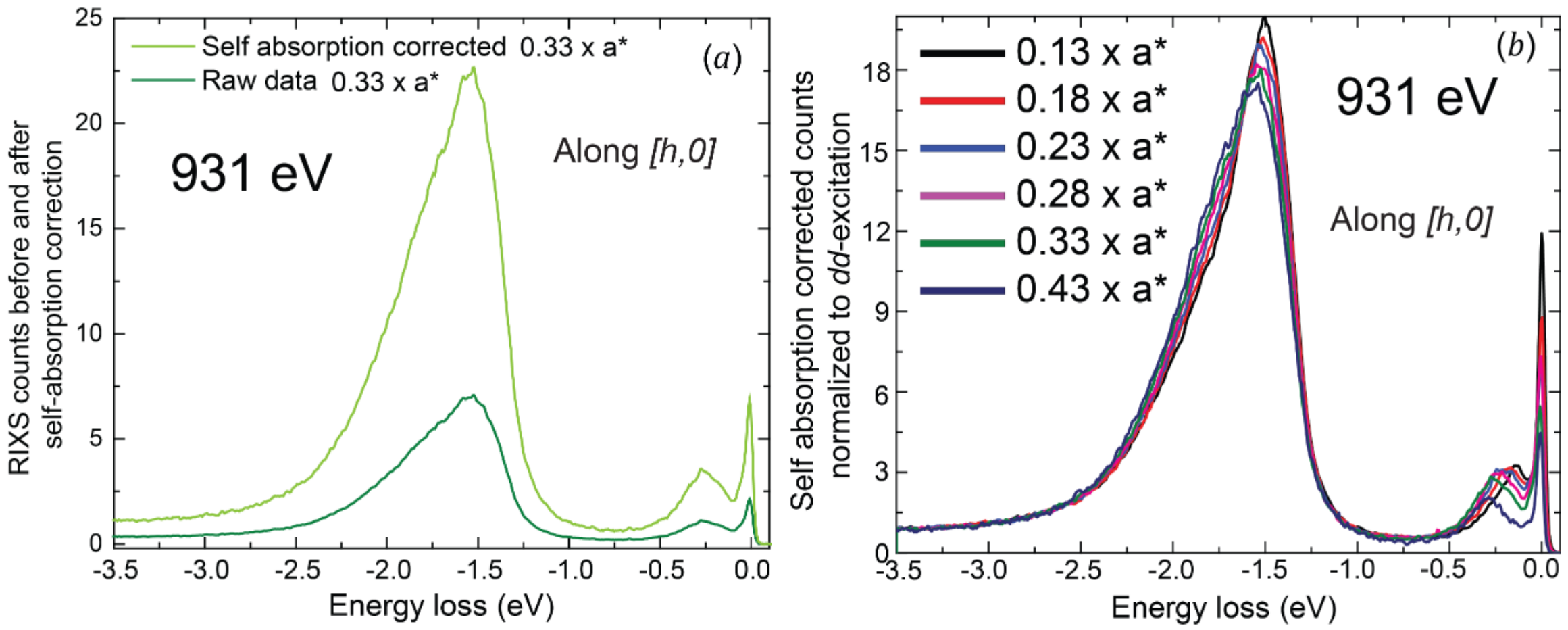

**Fig S5.** (a) Effect of self-absorption correction for a RIXS spectrum along [h,0] at 0.33a*. (b) Data normalization to the dd-excitation for the footprint correction of the RIXS spectra along [h,0] at 931 eV.

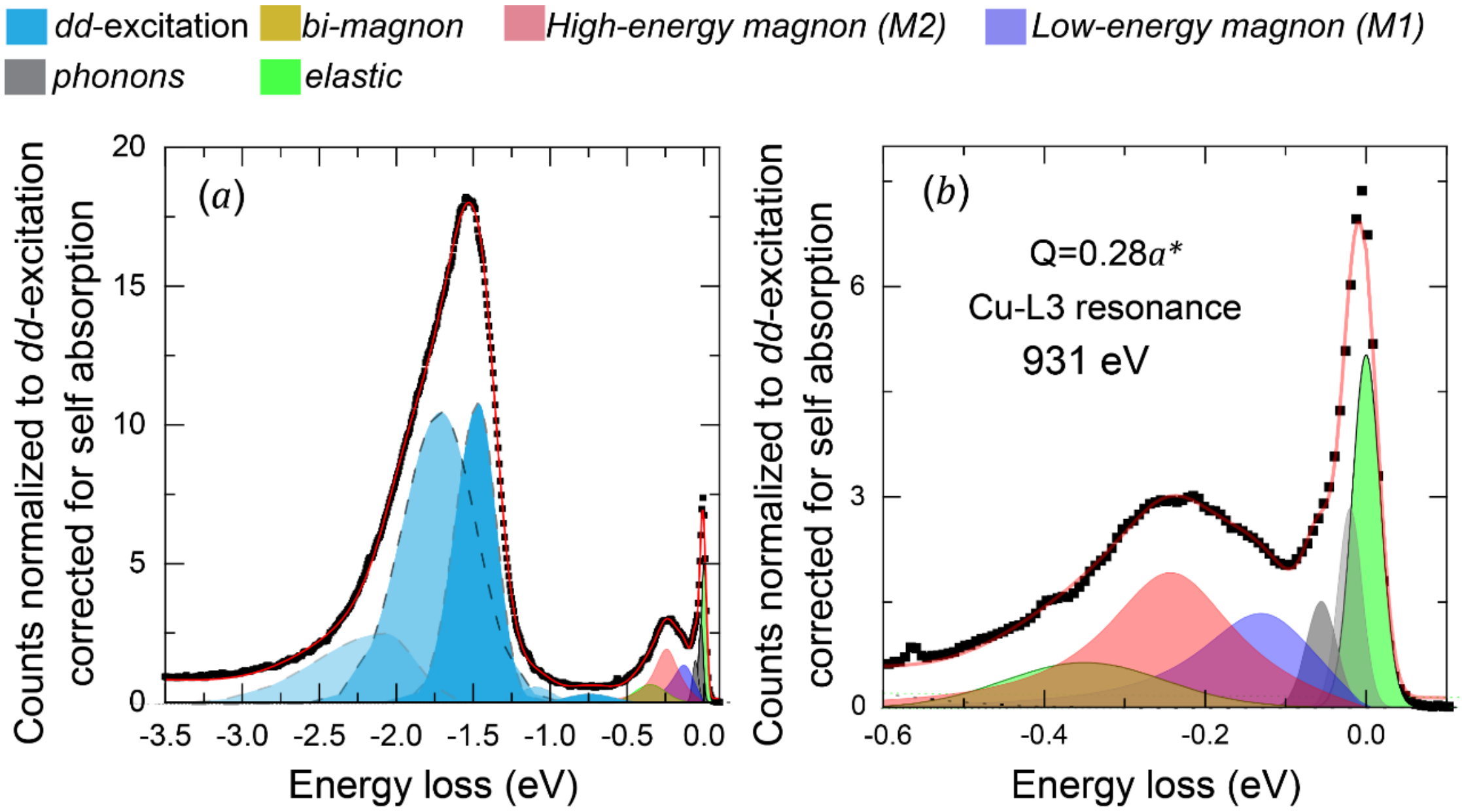

**Fig S6.** A model fit of a self-absorption corrected and normalized spectrum at 931 eV at 0.28a* along [h,0]. The various shaded and colored areas show the different contributions of the lattice, electronic and magnetic excitations.

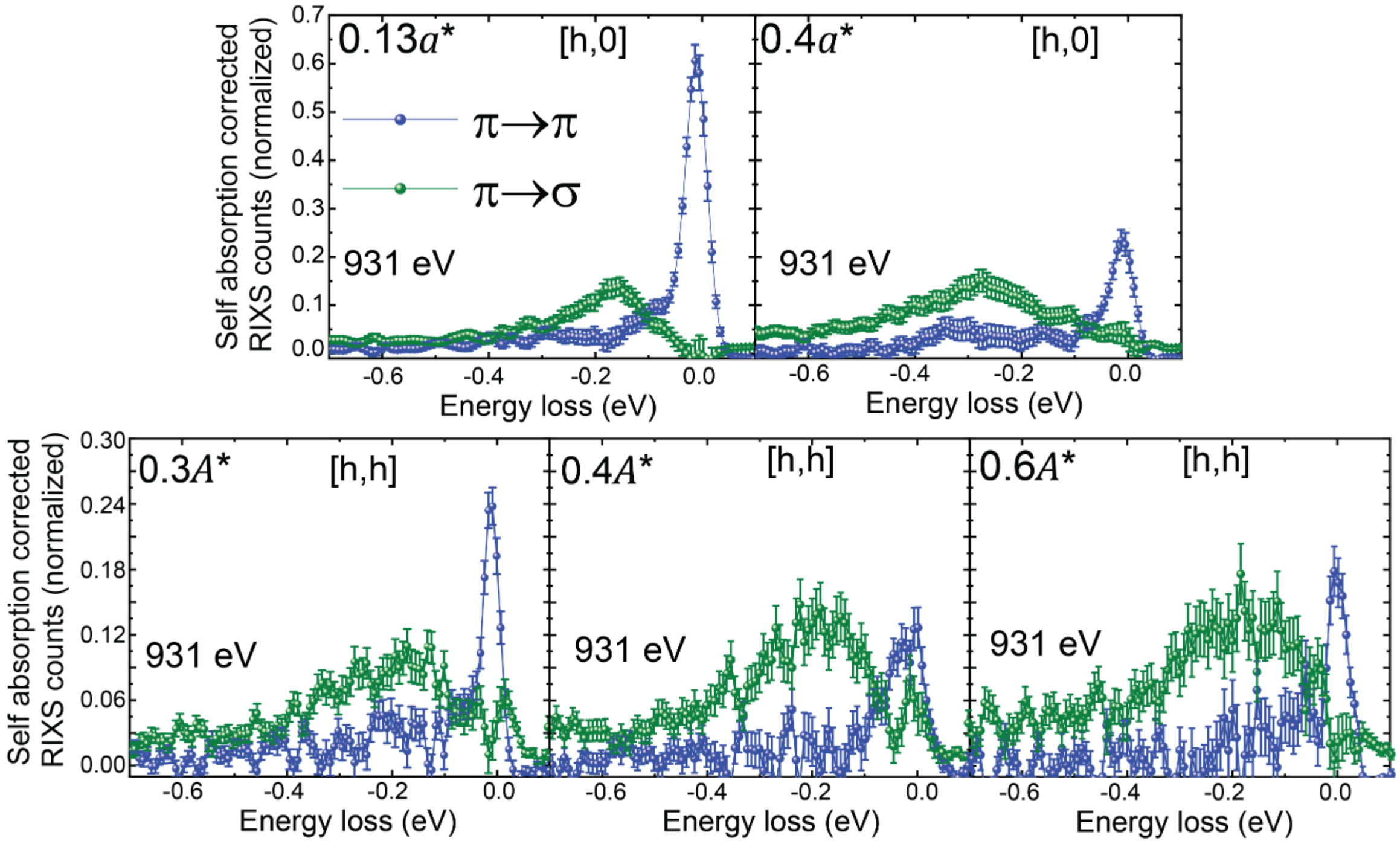

**Fig S7.** Results of all RIXS polarimetry analysis of the scattered beam: Note that the spin-flip component of the scattering dominates the neighborhood of M1 and M2 modes for all spectra. Directions and coordinates in the BZ have been indicated

.

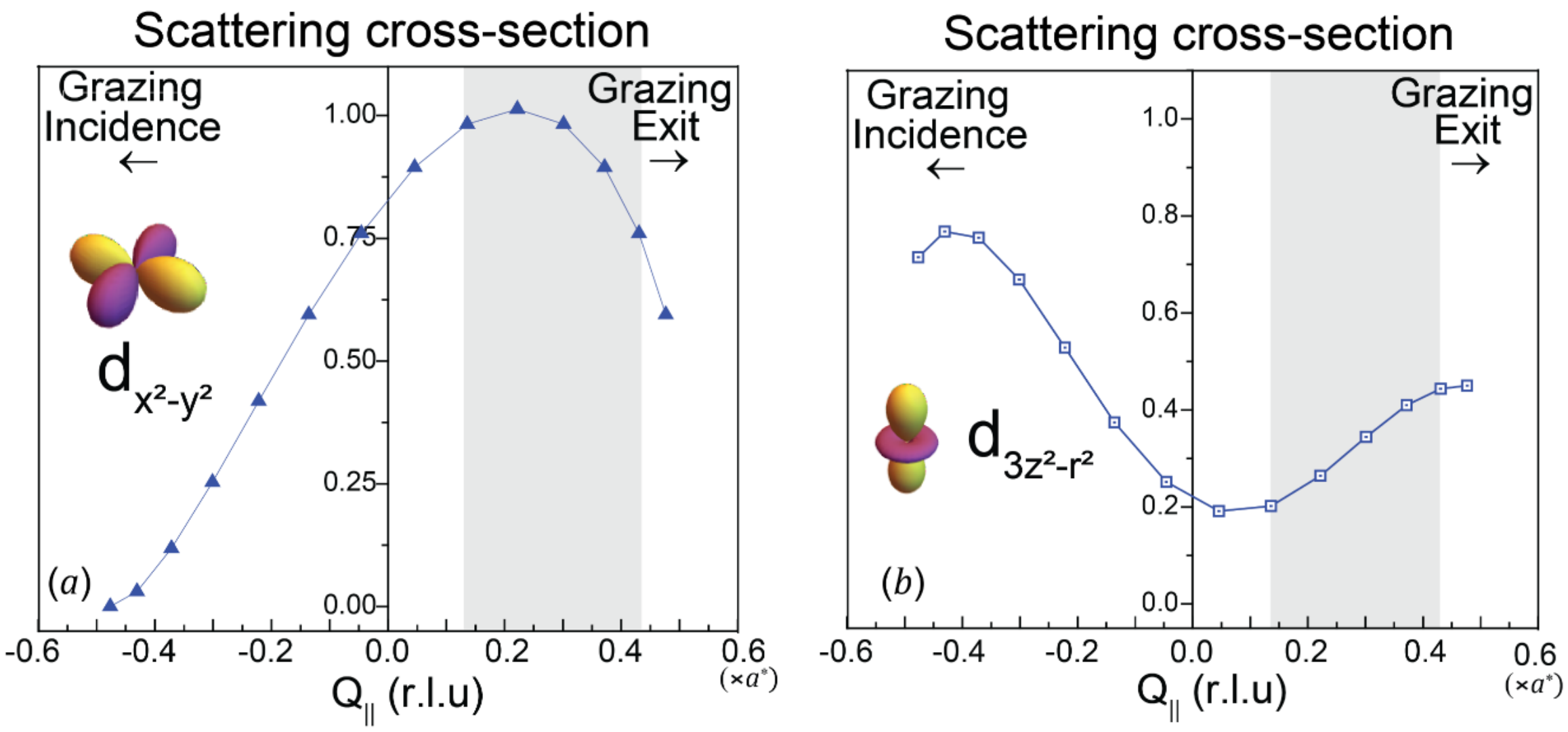

**Fig S8.** Summary of single ion model calculation for spin-flip scattering cross-section: (a) for $\boldsymbol{d_{x^2-y^2}}$ and (b) $\boldsymbol{d_{3z^2-r^2}}$ orbitals; the shaded region indicates the range of $\boldsymbol{Q}_{||}$ probed in our experiment along a*.